\documentclass[12pt]{article}
\usepackage[top=1.2in,bottom=1.2in, left = 1.2in, right = 1.2in]{geometry}
\usepackage[utf8x]{inputenc} 
\usepackage{amsmath,amssymb,amsfonts,amsthm,bbm}
\usepackage{epic,eepic,epsfig,longtable}
\usepackage{multirow,verbatim}
\usepackage{array}
\usepackage{lscape}
\usepackage{graphicx}
\usepackage{paralist}
\usepackage{latexsym}
\usepackage{comment}
\usepackage{booktabs}
\usepackage{epsfig}
\usepackage{setspace}
\usepackage{CJK}
\usepackage{color}
\usepackage{tikz}
\usepackage{hyperref}

\usepackage[authoryear,round]{natbib}
\makeatletter
\def\singlespace{\def\baselinestretch{1}\@normalsize}

\makeatletter
\def\singlespace{\def\baselinestretch{1}\@normalsize}

\renewcommand{\theequation}{\thesection.\arabic{equation}}
\numberwithin{equation}{section}
\renewcommand{\hat}{\widehat}
\renewcommand{\bar}{\overbar}

\renewcommand{\hat}{\widehat}

\newcommand{\bfm}[1]{\ensuremath{\mathbf{#1}}}

   \def\bA{\bfm A}  
   \def\bB{\bfm B}  
   \def\bC{\bfm C}  
     
\def\be{\bfm e}     
\def\bff{\bfm f}    \def\FF{\mathbb{F}}
   \def\bG{\bfm G}  
\def\bh{\bfm h}   \def\bH{\bfm H}  
   \def\bI{\bfm I}

   \def\bM{\bfm M}  
     
   \def\bO{\bfm O}  
     
\def\bq{\bfm q}

\def\bu{\bfm u}     
\def\bv{\bfm v}   \def\bV{\bfm V}  
\def\bw{\bfm w}     
\def\bx{\bfm x}   \def\bX{\bfm X}  
\def\by{\bfm y}

 \def\cT{{\cal  T}}

\newcommand{\0}{\mathbf{0}}
\newcommand{\1}{\mathbf{1}}

\newcommand{\B}[1]{\mbox{\normalfont\large\bfseries #1}}
\newcommand{\bfsym}[1]{\ensuremath{\boldsymbol{#1}}}

 \def\bmu{\bfsym {\mu}}                 
 
 \def\btheta{\bfsym {\theta}}           
 \def\beps{\bfsym \varepsilon}          \def\bepsilon{\bfsym \varepsilon}
              \def\bSigma{\bfsym \Sigma}
         
 \def\bomega {\bfsym {\omega}}

\def\bTheta{\bfsym{\Theta}}	
 \def\bvarpi{\bfsym {\varpi}}

\def\1{\bfsym{1}}	

	\newcommand{\bbE}{\mathbb{E}}

	\newcommand{\bbP}{\mathbb{P}}
	
	\newcommand{\bbR}{\mathbb{R}}

\DeclareMathOperator{\diag}{diag}

\DeclareMathOperator{\tr}{tr}
\newcommand{\T}{\top}
\newcommand{\ind}{\mathbbm{1}}

\def\eps{\varepsilon}
\def\beps{\mbox{\boldmath$\eps$}}

\def\today{\ifcase\month\or
  January\or February\or March\or April\or May\or June\or
  July\or August\or September\or October\or November\or December\fi
  \space\number\day, \number\year}

\newdimen\biblioindent    \biblioindent=30pt

 at 8truept

\def\eps{\varepsilon}

\newcommand{\beq}{\begin{equation}}
  \newcommand{\eeq}{\end{equation}}
\newcommand{\beqn}{\begin{eqnarray}}
  \newcommand{\eeqn}{\end{eqnarray}}
\newcommand{\beqnn}{\begin{eqnarray*}}
  \newcommand{\eeqnn}{\end{eqnarray*}}

\def \vec{\mathrm{vec}}
\def \Diag{\mathrm{Diag}}
\allowdisplaybreaks
\def\thesection{\arabic{section}}
\usepackage{verbatim}
\renewcommand{\baselinestretch}{1.66}
\def\tilde{\widetilde}

\def\FF{\mathcal{F}}
\def\[{\left [}  \def\]{\right ]} \def\({\left (}  \def\){\right )}
 \def\endpf{$\blacksquare$}
\newcommand{\overbar}[1]{\mkern 1.5mu\overline{\mkern-1.5mu#1\mkern-1.5mu}\mkern 1.5mu}
\def\hat{\widehat}

\newtheorem{thm}{Theorem}[section]

\newtheorem{lemma}{Lemma}[section]
\theoremstyle{remark}
\newtheorem{remark}{Remark}
\theoremstyle{proposition}

\newtheorem{assumption}{Assumption}

\newcommand{\thickhline}{%
    \noalign {\ifnum 0=`}\fi \hrule height 1pt
    \futurelet \reserved@a \@xhline
}
\newcolumntype{x}[1]{>{\centering\arraybackslash}p{#1}}

\usepackage{lipsum}

\title{Next Generation Models for Portfolio Risk Management: An Approach Using Financial Big Data}

 \author{Kwangmin Jung$^a$,  Donggyu Kim$^b$ and  Seunghyeon Yu$^b$\footnote{corresponding author. Tel: +82 2 958 3448.\newline
E-mail addresses: kwjung@postech.ac.kr (K. Jung),  donggyukim@kaist.ac.kr  (D. Kim),     ysh93@kaist.ac.kr  (S. Yu)} \\
\footnotesize{$^a$Department of Industrial and Management Engineering, POSTECH, Pohang, Korea}\\
\footnotesize{$^b$College of Business, Korea Advanced Institute of Science and Technology (KAIST), Seoul, Korea}}

\begin{document}
\maketitle
\begin{spacing}{1.5}

\begin{abstract}
This paper proposes a dynamic process of portfolio risk measurement to address potential information loss. The proposed model takes advantage of financial big data to incorporate out-of-target-portfolio information that may be missed when one considers the Value at Risk (VaR) measures only from certain assets of the portfolio. We investigate how the curse of dimensionality can be overcome in the use of financial big data and discuss where and when benefits occur from a large number of assets. In this regard, the proposed approach is the first to suggest the use of financial big data to improve the accuracy of risk analysis. We compare the proposed model with benchmark approaches and empirically show that the use of financial big data improves small portfolio risk analysis. Our findings are useful for portfolio managers and financial regulators, who may seek for an innovation to improve the accuracy of portfolio risk estimation.
\end{abstract}



\section{Introduction}


The widespread failure of portfolio risk management models observed in the 2008 financial crisis raises the question of how risk managers can analyze the risks of investment portfolios while taking into account systematic risks caused by a systemic crisis.
Extant literature has developed a range of risk measurement methods, for example, Value at Risk (VaR), expected shortfall (ES), entropic Value at Risk (EVaR), and superhedging price \citep{acerbi2002expected,ahmadi2012entropic,bensaid1992derivative,jorion2000value}.
VaR, in particular, has been widely used to estimate the level of risk by measuring the amount of loss for investments during a given period under normal market conditions \citep{jorion2000value}.
Especially, the literature documents the development of risk models with VaR to accommodate non-normal market conditions, for example, asymmetric, heavy-tailed stock returns, and volatility clustering (see, e.g., \citet{huisman1998var,natarajan2008incorporating,wu2007value,zoia2018value} for non-normal stock returns; \citet{billio2000value,broda2009chicago,engle2004caviar,fantazzini2008dynamic,mcneil2000estimation} for volatility clustering).

However, the 2008 financial crisis demonstrates that the development of risk models with dynamics for non-normal conditions does not guarantee financial firms to be successful in managing extreme systematic risks.
One of the supporting arguments can be that risk managers still consider only  stock data in their target portfolio \citep{fan2003semiparametric,hendricks1996evaluation,patton2019dynamic}.
They tend to overlook the risks of the assets that are not included in the portfolio; hence, the portfolio size for risk analysis can be relatively small.
This so-called small portfolio risk analysis may not sufficiently capture portfolio risk dynamics, particularly systematic risks inherent in the market.

In this regard, this study highlights a clear message that the use of financial big data can help risk managers take into consideration latent risks that cannot be captured in the portfolios of interest.
The contributions of this paper are threefold. 
First, we propose a dynamic risk model to resolve the above-mentioned problem by capturing systematic risk dynamics. 
At the core of the proposed model is the use of financial big data to analyze portfolio risks. 
The use of financial big data can help risk managers take into account a universe of assets  larger  than the size of their portfolios. 
This usefulness can lead to a higher accuracy of risk prediction, particularly, when it comes to capturing the systematic risk inherent in the markets. Consequentially, financial regulators can benefit from a high accuracy of risk prediction for market players to maintain the stability of financial markets and thus reduce the probability of a systemic risk occurrence.

One may be concerned about the fact that the use of financial big data can lead multivariate VaR methods to be computationally intensive in parameter estimation, which is called the curse of dimensionality.\footnote{For instance, when the BEKK model \citep{engle1995multivariate} is conducted on $p$-dimensional log return data, the number of parameters increases with the $p^2$ order. 
As the dimension $p$ increases, the parameter estimation becomes computationally demanding due to the exploding of computation time, which is the NP hard problem. 
Even if it is possible to estimate the parameters, they are not consistent estimators \citep{bickel2008covariance,engle2017large}.}
Our second contribution is to address this point in that the proposed model overcomes the curse of dimensionality with a two-step estimation.
The first step is to use the principal orthogonal complement thresholding (POET) procedure \citep{fan2013large} to estimate the latent factor and idiosyncratic component of the approximate latent factor model \citep{ait2017using, fan2018robust, fan2019robust, kim2018large, li2018embracing}.
The second step is a maximum quasi-likelihood estimation for the Generalized AutoRegressive Conditional Heteroskedasticity (GARCH) parameter with the latent factor estimator from the first step.
The estimated GARCH parameters are used to predict one-step ahead portfolio volatility, with which the one-step ahead VaR is predicted under the parametric and non-parametric $\sigma$-based approaches \citep{brooks2003volatility,giot2004modelling,hull1998incorporating}.

Lastly, our contribution is the incorporation of financial big data featuring the blessing of dimensionality \citep{donoho2000high, fan2019structured,fan2013large, li2018embracing} by discussing the asymptotic properties of $\sigma$-based approaches.
Upon the risk measurement procedure with integrated econometric approaches, our study is the first to suggest how a big data approach can help financial risk management. 
The proposed model can be particularly helpful for financial firms and regulators, who may be concerned about the stability of financial markets and question how big data can be optimally used for an improved  prediction of risks.

The rest of the paper is organized as follows. 
In Section \ref{sect:model}, we introduce the concept of VaR and develop a dynamic risk model.
In Section 3, we propose a maximum quasi-likelihood estimation method and study its asymptotic behavior.
Section \ref{sect:forecast} shows how the large volatility matrix can be predicted, how the proposed method can be applied to measure VaR, and how the curse of dimensionality can be solved.
In Section \ref{sect:Numerical}, we implement the Monte Carlo simulation to check the finite sample performance of the proposed methods and apply the proposed VaR measure to the empirical data.
We conclude in Section \ref{sect:conclusion}.\footnote{All technical proofs are provided in the online supplement (see Appendix A).}

\section{Model setup}     \label{sect:model}

Consider a market consisting of $p$ stocks.
Let $\by_t$ denote the $\bbR^p$-valued random vector of the assets' log returns at time $t$.
Then, the portfolio is characterized by a vector of asset weights $\bw\in \bbR^p$, for example, $r_t = \bw ^{\top} \by_t$. 
We assume that the log returns admit the factor model \citep{ait2017using,bai2003inferential,carhart1997persistence,fan2013large,fan2019robust,fama1992cross,fama1993common,fama2015five} as follows:  
\begin{align}    \label{eq:return}
    \by_{t} = \bmu_t+ \bV  \bff_t + \bu_{t},
\end{align}
where $\bmu_t$ is a mean vector,  $\bff_t$ is an $r$-dimensional factor, $\bV$ is a $p\times r$ factor loading matrix, and $\bu_t$ is an idiosyncratic component.
We assume that the factor $\bff_t$ and idiosyncratic component $\bu_t$ are independent, and  without loss of generality, we assume that the mean of $\bff_t$ and $\bu_t$ are zero.
Usually, the number of market factors is much smaller than the number of assets, so we assume that $r$ is finite. 
Moreover, in this paper, we consider the latent factor model \citep{ait2017using,fan2013large,fan2019robust,li2018embracing}, so $\bff_t$ and $\bu_t$ are not observable.  
In this section, under the latent factor model, we propose a VaR estimation procedure.

\subsection{$\sigma$-Based VaR method}     \label{sect:model:VaR}

One of the most popular risk measures is VaR, which estimates how much a portfolio of assets might lose under normal market conditions. 
Specifically, VaR at level $\alpha$ is defined as the $\alpha$-percentile of a portfolio log return distribution as follows:
\begin{equation}     \label{eq:VaR_def}
    \text{VaR}_{\alpha, t} =  - \inf{\big\{x : \mathbb{P}_{t-1} \{r_t \le x\} > \alpha \big\}},
\end{equation}
where $\mathbb{P}_{t-1}$ denotes the conditional distribution of the portfolio log returns $r_t$ given a filtration representing information to time $t-1$.
When the portfolio log return distribution follows a location-scale family, such as normal distribution and t-distribution, we can obtain VaR from the portfolio conditional mean $\mu_t$ and conditional volatility $\sigma_t$ given a filtration information to time $t-1$ as follows:
\begin{equation} \label{eq:VaR}
    \text{VaR}_{\alpha, t} =  -\mu_t - c_{\alpha} \sigma_t,	 
\end{equation}
where  $c_{\alpha}$ is an $\alpha$-quantile value of the distributions of the standardized $r_t$.
This method is called a $\sigma$-based approach \citep{jorion2000value}.
The performance of the  $\sigma$-based approach depends on  three components: the conditional mean $\mu_t$, conditional volatility $\sigma_t$, and quantile $c_{\alpha}$.
Quantile $c_{\alpha}$ can be determined using the parametric and non-parametric methods.
The parametric method involves the assumption of the parametric distribution for the standardized log return $(r_t - \mu_t) / \sigma_t$, such as normal distribution or student-t distribution, and $c_{\alpha}$ is determined based on the $\alpha$-quantile of each distribution.
The non-parametric method uses a sample quantile of the historical standardized portfolio log returns  $(r_t - \mu_t)/\sigma_t$.
Detailed implementations are described in Section \ref{sect:forecast:VaR}.
However, several empirical studies indicate that the portfolio conditional mean $\mu_t$ does not have strong time series patterns \citep{cajueiro2004hurst,fama1998market,malkiel2003efficient,narayan2004south}, which makes it difficult to predict the mean.
In this paper, we simply assume that the mean process is constant over time in order to reduce the model misspecification error.
Meanwhile, in the stock market, we observe volatility time series structures, such as  volatility clustering \citep{mandelbrot1963variation}. 
From this point of view, it is crucial to determine the dynamic structure of the conditional volatility $\sigma_t$.  
Thus, in this paper, we focus on how  the dynamic volatility structure can be incorporated in VaR.
We note that the mean structure would have a significant effect on VaR; hence, a parametric structure of the mean process should be developed.
We leave this for future study.

\subsection{Small portfolio risk analysis}     \label{sect:model:small}

When analyzing portfolio volatility, we usually investigate only the stock log return data that belong to the portfolio.
For example, let $\by_{t}^s$ be the $s$-dimensional log return vector whose stocks belong to the portfolio.
We also denote  the conditional co-volatility matrix of $\by_{t}^s$ by $\bSigma_{t}^s$ and  the $s$-dimensional portfolio weight vector by $\bw^s$.  
Then,  the portfolio conditional volatility can be obtained from $\bSigma_{t}^s$ as follows:
\begin{align*}
	\sigma_t = \sqrt{\bw^{s\T} \bSigma_{t}^s \bw^s}.
\end{align*}
To account for the market dynamics of the portfolio return, 
researchers have introduced several dynamic models, such as  the BEKK \citep{engle1995multivariate}, CCC \citep{bollerslev1990modelling}, and portfolio univariate GARCH \citep{bollerslev1986generalized}.
 See also \citet{engle2002dynamic}, \citet{kawakatsu2006matrix}, \citet{paolella2021non},  \citet{paolella2015comfort}, and \citet{van2002go}.
The BEKK model with the conditional mean vector $\bmu_{t}^s$ has the following structure:
\begin{align*} 
    \by_{t}^s &= \bmu_{t}^s + \(\bSigma_{t}^s\)^{\frac{1}{2}} \bepsilon_{t}^s,    \\
    \bSigma_{t}^s &= \bC \bC^\T + \bA^\T (\by_{t-1}^s-\bmu_{t-1}^s) (\by_{t-1}^s-\bmu_{t-1}^s)^\T \bA + \bB^\T \bSigma_{t-1}^s \bB, \\
    \varepsilon_{it}^s &\sim i.i.d.\;  F(0,1),
\end{align*}
where $\bC$ is an $s\times s$ lower triangular matrix, $\(\bSigma_{t}^s\)^{\frac{1}{2}}$ is  the square root of the matrix $\bSigma_{t}^s$,  $\bA$ and $\bB$ are $s\times s$ matrices, and $F(0,1)$ is some arbitrary distribution with mean zero and variance one.
As in the above BEKK model, the dynamic volatility models have the autoregressive form of the historical squared returns.
Empirical studies shows that the volatility is heterogeneous, and these models can explain the market dynamics \citep{bollerslev1986generalized, bollerslev1990modelling,  engle2002dynamic,engle1995multivariate, kawakatsu2006matrix,van2002go}.
Moreover, the dynamic volatility model helps to improve the measurement of VaR \citep{engle2004caviar,giot2004modelling,kuester2006value}. 
That is, the performance of measuring VaR can be improved by identifying the market dynamics.
In financial markets, we often observe that much of the market dynamic structure comes from the systematic risk, which is related to the entire  stock market; hence, the systematic risk information is contained in the entire stock. 
From this point of view, even if we consider only the small number of assets in the portfolio, to account for the market dynamics, we need to investigate the whole stock market and extract the systematic risk information from the stock returns.  
One of the naive extensions is to use a multivariate volatility model, such as the BEKK and CCC models, for the whole stock market.   
However, when applying these volatility models to the incorporation of a large number of assets, we face the curse of dimensionality \citep{bickel2008covariance, pakel2020fitting,engle2017large}.
Specifically, let $p$ be the total number of assets, and then the number of parameters quadratically increases with $p$.
Hence, with a large $p$, the estimated model is statistically inconsistent. 
Furthermore, in practice, the parameter optimization takes exponential computation time, which is known as the non-deterministic polynomial-time (NP) hard problem.
Thus,  applying  the usual multivariate volatility model to a large number of assets is not feasible. 
In this paper, we discuss how the curse of dimensionality issue can be handled while accounting for the market dynamics.

\subsection{High-dimensional dynamic volatility model }     \label{sect:model:factor}

In this section, we propose a dynamic factor model under the latent factor structure in \eqref{eq:return}. 
The factor model implies that the risk of assets stems from  factor and idiosyncratic components, which are called systematic and idiosyncratic risks, respectively. 
An idiosyncratic risk is related to the firm-specific risk; hence, it does not affect the entire market.
Moreover, it can be mitigated by the portfolio diversification. 
By contrast, a systematic risk arises from common market factor, such as interest rate, inflation, oil price, etc. 
Because systematic risk affects the entire market, we cannot mitigate this risk.
From this point of view, we impose a dynamic structure on the factor to account for the market risk. 
For the idiosyncratic risk, we employ some martingale assumption, for example, the idiosyncratic volatility is constant over time.
Then, the conditional expected co-volatility matrix of $\by_t$ given the current available information $\FF_{t}$ is expressed as follows:
\begin{align}    \label{eq:variance}
    \bbE \[ (\by_{t+1} - \bmu) (\by_{t+1} - \bmu) ^\T \middle | \FF_{t} \] =\bSigma_{t+1} = \bV \bSigma_{f,t+1}\bV^\T + \bSigma_{u} \text{ a.s.},
\end{align}
where $\bSigma_{f,t+1}= \bbE \[ \bff_{t+1} \bff_{t+1} ^{\top} \middle | \FF_{t} \] $, and $\bSigma_u$ represents the idiosyncratic volatility matrix. 
Under the framework \eqref{eq:variance}, the market dynamics can be explained by the factor volatility dynamics.

The latent factor model has an identification problem in $(\bV, \bff_t)$.
For example, the pair $(\bV, \bff_t)$ is not distinguishable from $(\bV \bH^\T, \bH \bff_t)$ for any orthonormal matrix $\bH$. 
To uniquely define the latent factor model, we often impose the following identifiability condition \citep{bai2012statistical,bai2013principal, fan2013large,kim2019factor,li2018embracing}: 
\begin{align}    
     \bV^\T \bV = p\bI_r \quad   \text{ and }  \quad 
     \bSigma_{f,t}\text{ is diagonal}.    	\label{eq:identifiability}
\end{align} 
The identification condition \eqref{eq:identifiability} implies that the scaled factor loading matrix $p^{-1/2}\bV$ and  elements of $p  \bSigma_{f,t}$ are the eigenmatrix and eigenvalues of the conditional variance of the factor part.
Thus, the market dynamics are explained by the dynamic structure of the eigenvalues, whereas its factor loading matrix is constant over time.

To account for the factor dynamics, we employ the famous GARCH model \citep{bauwens2006multivariate,engle1995multivariate,engle2002dynamic,engle2016dynamic,van2002go}.
The conditional expected volatility of the factor is modeled by the following GARCH structure:
\begin{align}       \label{eq:f_garch}
	&\bff_t =   \( \sqrt{ h_{1t} (\btheta)} \epsilon_{1t}, \ldots, \sqrt{ h_{rt}(\btheta)}\epsilon_{rt}  \) , \notag \\
    &\bh_t (\btheta)=  \bomega + \bA \bff_{t-1}^2 + \bB \bh_{t-1} (\btheta),
\end{align}
where $\epsilon_{it}$'s are $\text{i.i.d.}$ random variables with mean zero and unit variance;  $\bh_{t}(\btheta)$ is an $r$-dimensional vector of the conditional expected factor volatility $\bSigma_{f,t}$, that is, $\bh_{t}= \big(h_{1t} (\btheta), \ldots,\allowbreak h_{rt} (\btheta)\big)   = \diag(\bSigma_{f,t})$, where $\diag(\bX)$ is a vector whose elements are diagonal entries of $\bX$, $\bff_t^2$ is an element-wise squared vector of the factor log return $\bff_t$, and $\btheta =( \bomega, \vec(\bA), \vec(\bB))$ is the GARCH parameters. 
The GARCH model in \eqref{eq:f_garch} indicates that the conditional expected volatility of the factor is the autoregressive form of the historical squared factor log returns. 
Thus, we expect that this model can capture the stylized market features, such as the volatility clustering and heavy tail.
The empirical study in Section \ref{sect:empirics}  also supports this.

The idiosyncratic component comes  from the firm-specific risk, so their risk is not strongly connected.  
Empirical studies show that there are some local factors that affect few other idiosyncratic components \citep{ait2017using, boivin2006more}. 
In light of these, we allow a weak relationship between idiosyncratic components so that the idiosyncratic volatility $\bSigma_u =\( (\Sigma_{u,ij})_{i,j=1,\ldots, p} \)$ satisfies the following sparse condition:
\begin{align}   \label{eq:sparsity}
    \max_{i\le p}{\sum_{j=1}^p{|\Sigma_{u,ij}|^q}(\Sigma_{u, ii}\Sigma_{u, jj})^{(1-q)/2}} &\le s_p,
\end{align}
where  $ q \in [0, 1)$,  and the sparsity measure $s_p$ diverges slowly with the dimension $p$, for example, $\log{p}$.
We define $0^0 =0$. 
This sparsity condition is widely employed in the large convariance matrix inferences \citep{ait2017using,bickel2008covariance, cai2011adaptive, fan2013large, fan2019robust, kim2019factor}.
When the idiosyncratic volatility satisfies the exact sparsity, that is,   $q=0$,  the sparsity condition indicates that each asset has at most $s_p$ non-zero idiosyncratic correlations with other assets.

Under the volatility structure \eqref{eq:variance} and \eqref{eq:f_garch},  the VaR of portfolios in \eqref{eq:VaR} can be calculated as follows:
\begin{align*} 
    \text{VaR}_{\alpha, t} =  -\bw^\T \bmu - c_{\alpha} \sqrt{\bw^\T (\bV \bSigma_{f,t}\bV^\T + \bSigma_{u}) \bw}.
\end{align*}
To evaluate the above VaR value, we need to estimate the unobserved factor components and the idiosyncratic volatility.
However, to incorporate the entire market asset information, we consider a large number of assets and consequently run into the curse of dimensionality problem. 
In the following section, we discuss how to overcome this issue and introduce an estimation procedure to incorporate the financial big data for risk analysis.

\section{Factor dynamics estimation}     \label{sect:estimation}

First, we define the notations.
We denote $\|\cdot\|,\|\cdot\|_F$, and $\|\cdot\|_{\max}$ by the matrix spectral norm, Frobenius norm, and max norm, respectively.
We use $O_p$ as a big-O in probability, $\lambda_k(\bA)$ as the $k^{th}$ largest eigenvalue of the square matrix $\bA$, and $\vec(\bA)$ as the vectorization of $\bA$. 
We also denote the true parameters by $\btheta_0 = (\bomega_0, \vec (\bA_0), \vec(\bB_0) )$.

\subsection{Latent factor estimation}    \label{sect:estimation:factor}

To estimate the model parameter $\btheta_0 = (\bomega_0, \vec (\bA_0), \vec(\bB_0) )$, we first need to estimate the latent factor components $\bff_t$ and $\bV$.  
We recall that the conditional volatility matrix is 
\begin{align*}  
    \bSigma_{t+1} = \bV \bSigma_{f,t+1}\bV^\T + \bSigma_{u},
\end{align*}
and  the mean conditional volatility matrix is
\begin{equation*}
	\bar{\bSigma} = \bV \bar{\bSigma}_{f} \bV^T + \bSigma_{u},
\end{equation*}
where $\bar{\bSigma}_{f} = T^{-1} \sum_{t=1}^T \bSigma_{f,t}$. 
Under the identification and sparsity conditions \eqref{eq:variance} and \eqref{eq:sparsity}, the mean conditional volatility matrix has the low-rank plus sparse structure that is widely used in the high-dimensional factor analysis \citep{ait2017using,fan2013large, fan2019robust,kim2019factor}. 
With the low-rank plus sparse structure,  \citet{fan2013large} introduced the POET estimation procedure to estimate the latent factor volatility and idiosyncratic volatility matrices.
To harness the POET procedure, we need a good proxy of the mean conditional volatility matrix.   
We use the sample covariance matrix $\hat{\bar{\bSigma}} = T^{-1} \sum_{t=1}^T (\by_t  - \bar{\by} ) (\by_t - \bar{\by}) ^\T$ with sample mean $\bar{\by}=T^{-1}\sum_{t=1}^T{\by_t}$, and under some mild condition,  the martingale convergence theorem implies that the sample covariance matrix converges to $\bar{\bSigma}$.
Thus, we apply the POET procedure with the sample covariance matrix to estimate the latent factor components. 
Specifically, the eigenvalue decomposition admits 
\begin{align}     \label{eq:spectral}
    \hat{\bar{\bSigma}}  = \sum_{i=1}^p {\hat{\lambda}_i \hat{\bq}_i {\hat{\bq}_i}^\T},
\end{align}
where $\hat{\lambda}_k$ and $\hat{\bq}_k$ are the $k^{th}$ largest eigenvalues and eigenvectors of the sample covariance matrix $\hat{\bar{\bSigma}}$, respectively. 
Then, the factor loading matrix estimator $\hat{\bV}$ is $\sqrt{p}( \hat{\bq}_1, \ldots, \hat{\bq}_r)$, and the mean factor volatility matrix estimator $\hat{\bar{\bSigma}}_{f}$ is $p^{-1}\Diag \big( (\hat{\lambda}_1, \ldots, \hat{\lambda}_r )\big)$, where $\Diag(\bx)$ is a diagonal matrix whose diagonal entries are $\bx$.
Note that $\hat{\bV}$ has a multiplier $\sqrt{p}$ so that $\hat{\bV}$ satisfies the condition $\bV^\T\bV = p\bI_r$.
Then, the latent factors can be obtained as follows: 
\begin{align}\label{f-est}
   \hat{\bff}_t =  \frac{1}{p}\hat{\bV}^\T (\by_{t} - \bar{\by}).   
\end{align}
 
Theorem \ref{thm:factor} in the online supplement shows that  the convergence rate of  $\hat{\bff}_t^2$ is $1/\sqrt{T}  + \sqrt{s_p/p}$. 
 We note that when the factor is not observable and the number of assets is finite,  the latent factors are impossible to estimate. 
By contrast, Theorem \ref{thm:factor} shows the blessing of financial big data in the latent factor estimation.
Specifically, every asset contains the common market information.
Thus, as the number of assets increases, the information for the latent factors also increases.    
Therefore, more stock samples exhibit a clearer latent signal, and we can estimate the latent factors consistently with the convergence rate $\sqrt{s_p/p}$.

\subsection{Maximum quasi-likelihood estimation}     \label{sect:estimation:GARCH}

In this section, we propose a model parameter estimation procedure. 
Specifically, we adopt the following quasi-maximum likelihood estimation (QMLE) method:
\begin{align}     \label{eq:quasilikelihood_true}
     \min_{\btheta \in \bTheta}{\sum_{t=1}^T{\sum_{i=1}^r{\left(\log{h_{it}(\btheta)}  +  h_{it}^{-1}(\btheta)f_{it}^2\right)}}},
\end{align}
where $\bTheta$ is a compact parameter space.
Then, the well-developed asymptotic theorems for the QMLE provide the consistency \citep{comte2003asymptotic}.  
However, the factors are not observable; hence, to evaluate the quasi-likelihood function, we use the non-parametric factor estimator $\hat{\bff_t^2}$. 
For example, the conditional co-volatility $\bh_t(\btheta)$ is estimated by
\begin{align}     
    \hat{\bh}_t(\btheta)   =  \bomega + \bA \hat{\bff}_{t-1}^2 + \bB \hat{\bh}_{t-1}(\btheta),    \label{eq:GARCH_est}  
\end{align}
and we use  $\hat{\bh}_1(\btheta) =  (\bI_r - \bA - \bB)^{-1} \bomega$ as the initial value. 
Note that $\hat{\bh}_1(\btheta)$ is the unconditional volatility of the factor. 
Then, we calculate the quasi-likelihood function with the conditional co-volatility estimator $\hat{\bh}_t(\btheta)$ and obtain the maximum quasi-likelihood estimator $\hat{\btheta}$  as follows:
\begin{align}     \label{eq:quasilikelihood}
    \hat{\btheta}   =  \arg\min_{\btheta \in \bTheta}{\sum_{t=1}^T{\sum_{i=1}^r{\left(\log{\hat{h}_{it}(\btheta)}  +  \hat{h}_{it}^{-1}(\btheta)\hat{f}_{it}^2\right)}}}.
\end{align}
Theorem \ref{thm:GARCH} in the online supplement shows the consistency of $\hat{\btheta}$.

%

\section{Large volatility matrix estimation and VaR forecast}     \label{sect:forecast}

\subsection{Estimation of the one-step ahead large volatility matrix} 	\label{sect:forecast:volatility}

In the previous section, we find the blessing of dimensionality in the latent factor estimation.
However, when estimating large volatility matrices, we still suffer from the curse of dimensionality. 
For example, sample volatility matrix estimators are inconsistent when both the number of assets and sample size go to infinity \citep{bickel2008covariance, cai2011adaptive}.
To overcome the curse of dimensionality, we impose the  sparse structure \eqref{eq:sparsity} on the idiosyncratic volatility matrix.
As discussed in Section \ref{sect:model:factor}, in the stock market, the co-movement of stocks can be explained by the common factor, and the remaining idiosyncratic co-volatilities are weakly correlated. 
 Thus, the sparsity condition is realistic. 
To estimate the sparse idiosyncratic volatility matrix, we employ  the POET procedure \citep{fan2013large} as follows.
First, we estimate the input idiosyncratic volatility matrix estimator by using the non-pervasive eigen-components as  follows:
$$
	\hat{\bSigma}_u = \sum_{i=r+1}^p {\hat{\lambda}_i \hat{\bq}_i {\hat{\bq}_i}^\T},
$$
where the  eigenvalues $\hat{\lambda}_i$'s and eigenvectors $\hat{\bq}_i$'s are defined in \eqref{eq:spectral}.
Then, we apply the thresholding method to the input idiosyncratic volatility matrix estimator as follows:
\begin{align}        \label{eq:thresholding}
    [\cT(\hat{\bSigma}_u)]_{ij} = \begin{cases} \hat{\Sigma}_{u,ii}, &\text{ if }i=j \\  
s_{ij}(\hat{\Sigma}_{u,ij})\ind_{\big(|\hat{\Sigma}_{u,ij} | \ge \tau_T\sqrt{ \hat{\Sigma}_{u,ii}  \hat{\Sigma}_{u,jj}}\big)}, &\text{ if } i\neq j \end{cases},    
\end{align}
where $\ind_{(\cdot)}$ is an indicator function, and $s_{ij}(\cdot)$ is a shrinkage function satisfying $|s_{ij}(x) - x| \le \tau_{T} \sqrt{\hat{\Sigma}_{u,ii}  \hat{\Sigma}_{u,jj}}$. 
The examples of the shrink function $s_{ij}(x)$ are the soft thresholding function $s_{ij} (x) = x -  sign(x)\tau_T \sqrt{ \hat{\Sigma}_{u,ii}  \hat{\Sigma}_{u,jj}}$ and the hard thresholding function $s_{ij} (x) = x$.
The thresholding level $\tau_T$ will be given in Theorem \ref{thm:prediction}.
The working principle of the thresholding method is that the co-volatility is zero if the estimated correlation is weak.
This makes the estimated idiosyncratic volatility matrix estimator sparse, so the estimated idiosyncratic volatility satisfies the sparse condition.

 With the estimated factor volatility $\hat{\bSigma}_{f,t+1}$ and idiosyncratic volatility matrix $\cT(\hat{\bSigma}_u)$, we estimate the conditional volatility matrix as follows:
\begin{align}     \label{eq:volatiliy_prediction}
 \hat{\bSigma}_{t+1} = \hat{\bV} \hat{\bSigma}_{f,t+1} \hat{\bV}^\T + \cT(\hat{\bSigma}_u). 
\end{align}
We call the conditional volatility matrix estimator $ \hat{\bSigma}_{t+1}$  the P-GARCH estimator.

The following theorem shows the asymptotic behaviors for the P-GARCH estimator.
\begin{thm}     \label{thm:prediction} 
Under  the model in Section \ref{sect:model:factor}, suppose that Assumptions \ref{ass:factor}--\ref{ass:prediction} in the online supplement hold. 
Take the thresholding level as $\tau_T = C_{\tau} \big(\sqrt{\log{p}/T} + \sqrt{s_p/p} \big)$ for some positive constant $C_{\tau}$. 
Then, we have
\begin{align*}     
    &\left\|\cT(\hat{\bSigma}_{u}) - \bSigma_{u} \right\|_{\max} = O_p\left(\tau_T\right),   \quad \left\|\cT(\hat{\bSigma}_{u}) - \bSigma_{u} \right\| = O_p\left(s_p \tau_T^{1-q}\right),    \\
    &\left\|\hat{\bSigma}_{t+1} - \bSigma_{t+1} \right\|_{\max} = O_p\left(\tau_T\right),  \quad \left\|\hat{\bSigma}_{t+1} - \bSigma_{t+1} \right\|_{\bSigma_{t+1}} = O_p\left(\frac{\sqrt{p}}{T} + s_p \tau_T^{1-q}\right),
\end{align*}  
where the relative Frobenius norm $\|\bG_1\|_{\bG_2}^ 2 = p^{-1} \| \bG_2^{-\frac{1}{2}} \bG_1 \bG_2^{-\frac{1}{2}} \|_F^2$ for any given $p\times p$ matrices $\bG_1$ and $\bG_2$.
\end{thm}

\begin{remark}
Theorem \ref{thm:prediction}  shows that the P-GARCH is the consistent estimator as long as $p=o(T^2)$.
\end{remark}

With the realistic low-rank plus sparse structure, we can enjoy the blessing of  dimensionality  for estimating the factor volatility matrix.
Moreover, using the regularization method, as shown in Theorem \ref{thm:prediction}, we can overcome the curse of dimensionality.

\subsection{One-step ahead VaR forecast from financial big data} 	\label{sect:forecast:VaR}

In this section, we discuss how   the VaR value can be measured with the P-GARCH estimator $\hat{\bSigma}_{t+1}$ in \eqref{eq:volatiliy_prediction}. 
Using the plug-in method, we estimate the one-step ahead VaR  as follows:
\begin{align}     \label{eq:VaR_est}
    \hat{\text{VaR}}_{\alpha, t+1} = -\bw^\T \bar{\by}  - c_{\alpha} \sqrt{\bw^\T \hat{\bSigma}_{t+1}\bw},
\end{align}
where $c_\alpha$ is an $\alpha$-quantile, and $\bar{\by}$ is a sample mean vector.
To evaluate the VaR value, we need to determine the $\alpha$-quantile value. 
To do this, we assume that the standardized portfolio log returns are i.i.d. 
Then, when the standardized portfolio log returns follow the standard normal distribution, $c_{\alpha}$ is the $\alpha$-quantile of the standard normal $z_{\alpha}$.
When the standardized portfolio log returns follow the multivariate t-distribution with the degrees of freedom $\nu$, then $c_{\alpha}=t_{\nu,\alpha}\sqrt{(\nu-2)/\nu}$, where $t_{\nu,\alpha}$ is an $\alpha$-quantile of the t-distribution with  the degrees of freedom $\nu$ \citep{glasserman2002portfolio}.
We call them the parametric $\sigma$-based VaR estimator.
The performance of the parametric $\sigma$-based VaR estimator depends on the distribution assumption.
On the other hand, to obtain distribution robust VaR estimators, we use the non-parametric sample quantile method, which we call the non-parametric $\sigma$-based VaR estimator.
For example,  $c_{\alpha}$ is set to be $\lceil \alpha T \rceil$-th smallest value of  $\{(\bw^\T (\by_t - \bar{\by})/ (\bw^\T \hat{\bSigma}_{t}\bw)^{1/2}\}_{t=1}^T$, where $\lceil \cdot \rceil$ is a ceiling function.

Then, the following theorem shows the convergence rates of VaR.

\begin{thm}     \label{thm:prediction_VaR} 
Under the model in Section \ref{sect:model:factor}, suppose that Assumptions \ref{ass:factor}--\ref{ass:prediction_VaR:subgaussian} in the online supplement hold. 
Then, for an arbitrary portfolio weight $\bw$ with the gross exposure constraint $\|\bw\|_1 \le C$, the parametric $\sigma$-based VaR estimator has
\begin{align*}
    &\left | \hat{\text{VaR}}_{\alpha,t+1} - \text{VaR}_{\alpha,t+1}\right | = O_p\left(\sqrt{\frac{\log{p}}{T}} + \sqrt{\frac{s_p}{p}}\right).
\end{align*}
Moreover,  the non-parametric $\sigma$-based VaR estimator is
\begin{align*}
    \left | \hat{\text{VaR}}_{\alpha,t+1} - \text{VaR}_{\alpha,t+1}\right | = O_p\left( \sqrt{ \log{T}}\left(\sqrt{\frac{\log{p}}{T}} + \sqrt{\frac{s_p}{p}}\right)\right).
\end{align*}
\end{thm}


 In this paper, we focus on investigating the effect of the VaR estimator with the financial big data for a small portfolio.
When we do not incorporate financial big data, that is, $p$ is finite, the absolute VaR error is not consistent for all parametric and non-parametric estimators.
This is because the term $\sqrt{s_p/p}$ does not converge and so the latent factor estimation error is dominant. 
Therefore, Theorem \ref{thm:prediction_VaR} supports that incorporating financial big data leads to the blessing in the VaR forecast by capturing common factor dynamics.

\begin{remark}
In this paper, we consider the parametric and non-parametric quantile structures.
However, imposing the parametric structure, such as normal and Student $t$ distributions, on the standardized returns does not explain the skewness of the asset returns. 
Furthermore, the empirical quantile approach can have large errors in the extreme tail and does not work well when we calculate other risk meausures, such as expected shortfall. 
Thus, it is important and interesting to develop a quantile estimation procedure, which is able to account for  stylized features of the asset returns, such as skewness and heavy-tailedness.  
We leave this for future study.
\end{remark}

\section{Numerical studies} 	\label{sect:Numerical}

\subsection{Simulation study} 	\label{sect:Monte}

In this section, we conducted Monte Carlo simulations to check the finite sample performances of the proposed P-GARCH and corresponding VaR model.
 The data-generating process is analogous to the models \eqref{eq:return} and \eqref{eq:variance}.
The number of factors was chosen to be $r=3$; the mean of $\by_t$ is set to be zero $\bmu=\0$; and the factor loading matrix $\bV$ was randomly sampled from the first $r$ right singular vector of random matrix, which has the elements $\text{i.i.d.}\,\text{Unif}(0,1)$. 
The latent factors $\bff_t$ were generated by the multivariate normal distribution with the conditional expected volatility $\bSigma_{f,t}$ as follows:
\begin{align*}
	\bSigma_{f,t}  = \Diag(\bh_{t} (\btheta_0)) \quad  \text{ and } \quad  	\bh_{t}(\btheta_0)  = \bomega_0 + \bA_0 \bff_{t-1}^2 + \bB_0 \bh_{t-1}(\btheta_0),  
\end{align*}
where 	$\bh_{1}(\btheta_0)  = \left(\bI_r - \bA_0 -\bB_0 \right)^{-1} \bomega_0$,
\begin{align*}
	\bomega_0 = \begin{pmatrix}
	0.003 	\\
	0.002 	\\
	0.001 	\\
	\end{pmatrix},\qquad
	\bA_0 = \begin{pmatrix}
	0.2 & 0.3 & 0.4 	\\
	0.15 & 0.12 & 0.2 	\\
	0.1 & 0.1 & 0.1 	
	\end{pmatrix}, \qquad
	\bB_0 = \begin{pmatrix}
	0.2 & 0.1 & 0.1 	\\
	0.2 & 0.05 & 0.07 	\\
	0.1 & 0 & 0.05
	\end{pmatrix}.
\end{align*}
The idiosyncratic risk $\bu_t$ was generated from the multivariate normal distribution with the following sparse co-volatility:
\begin{align} 	\label{eq:banded_sigma_u}
	\Sigma_{u,ij} = 0.01\times 0.5^{|i-j|}.
\end{align}
We varied $p$ from 20 to 500 and $T$  from 500 to 10,000, and employed the QMLE procedure in Section \ref{sect:estimation:GARCH} to estimate the GARCH parameters.
We repeated the entire procedure 500 times.

\begin{table}[h!]
\caption{Mean absolute error (MAE) of the QMLE estimate $\hat{\btheta}$ with 500 replications.}
\vspace{2mm}
\label{table:MAE1}
{\centering
\renewcommand\arraystretch{1.1}
\begin{tabular}{rrrrrrrrrrr}  
\thickhline
\multicolumn{1}{c}{\multirow{2}{*}{$p$}}  & \multicolumn{1}{c}{\multirow{2}{*}{$T$}}  &\multicolumn{9}{c}{MAE$\times 10^2$} \\ \cline{3-11}
                &   & \multicolumn{1}{c}{$\omega_1$} & \multicolumn{1}{c}{$\omega_2$} & \multicolumn{1}{c}{$\omega_3$} & \multicolumn{1}{c}{$A_{11}$} & \multicolumn{1}{c}{$A_{12}$} & \multicolumn{1}{c}{$A_{13}$} & \multicolumn{1}{c}{$B_{11}$} & \multicolumn{1}{c}{$B_{12}$} & \multicolumn{1}{c}{$B_{13}$} \\ 
\hline
\multirow{5}{*}{20}  & 500   & 0.138 & 0.088 & 0.068 & 5.706 & 9.483 & 12.637 & 14.279 & 18.768 & 21.801 \\
    & 2000  & 0.060 & 0.047 & 0.042 & 2.816 & 4.343 &  6.830 & 11.338 & 13.448 & 15.169 \\
    & 4000  & 0.046 & 0.040 & 0.034 & 2.033 & 3.359 &  5.621 &  9.738 & 10.703 & 14.345 \\
    & 6000  & 0.039 & 0.032 & 0.031 & 1.736 & 2.887 &  5.436 &  8.871 &  9.450 & 12.538 \\
    & 10000 & 0.030 & 0.031 & 0.029 & 1.333 & 2.825 &  4.947 &  8.279 &  8.912 & 12.923 \\
[.4cm]
\multirow{5}{*}{100} & 500   & 0.124 & 0.079 & 0.060 & 5.816 & 8.651 & 12.264 & 13.739 & 16.993 & 19.134 \\
    & 2000  & 0.056 & 0.040 & 0.028 & 2.594 & 4.306 &  5.933 & 10.340 & 12.012 & 13.912 \\
    & 4000  & 0.040 & 0.027 & 0.020 & 1.846 & 2.925 &  3.978 &  9.869 & 10.098 & 12.730 \\
    & 6000  & 0.029 & 0.021 & 0.015 & 1.550 & 2.263 &  3.147 &  8.539 &  8.335 & 11.924 \\
    & 10000 & 0.023 & 0.018 & 0.012 & 1.047 & 1.874 &  2.606 &  7.993 &  7.731 & 10.932 \\
[.4cm]
\multirow{5}{*}{500} & 500   & 0.118 & 0.079 & 0.055 & 5.788 & 8.539 & 11.931 & 14.059 & 16.716 & 18.482 \\
    & 2000  & 0.058 & 0.039 & 0.027 & 2.734 & 4.208 &  5.619 & 10.161 & 11.007 & 13.436 \\
    & 4000  & 0.039 & 0.027 & 0.019 & 1.786 & 2.916 &  3.948 &  8.975 &  9.451 & 12.525 \\
    & 6000  & 0.031 & 0.023 & 0.015 & 1.546 & 2.448 &  3.571 &  8.615 &  8.533 & 11.790 \\
    & 10000 & 0.024 & 0.018 & 0.012 & 1.057 & 1.764 &  2.456 &  7.932 &  7.477 & 10.843 \\    
\thickhline
\end{tabular}
}
{Note: To save space, we report only the first 3 elements of $\hat{\bA}$ and $\hat{\bB}$ are reported. 
The other results have a similar pattern.}
\end{table}

Table \ref{table:MAE1} shows the mean absolute error (MAE) of the QMLE estimate $\hat{\btheta}$ for various $p$ and $T$. 
For the sake of spatial efficiency, we only documented estimation results with nine parameters.\footnote{The estimations with the rest of parameters provide similar results, which are available upon request.}
The result illustrates that MAEs decreases as the parameter as $T$ or $p$ increases, which  supports the theoretical  findings in Theorem \ref{thm:GARCH}. 

\begin{figure}[h!]
\centering
\includegraphics[width = \textwidth]{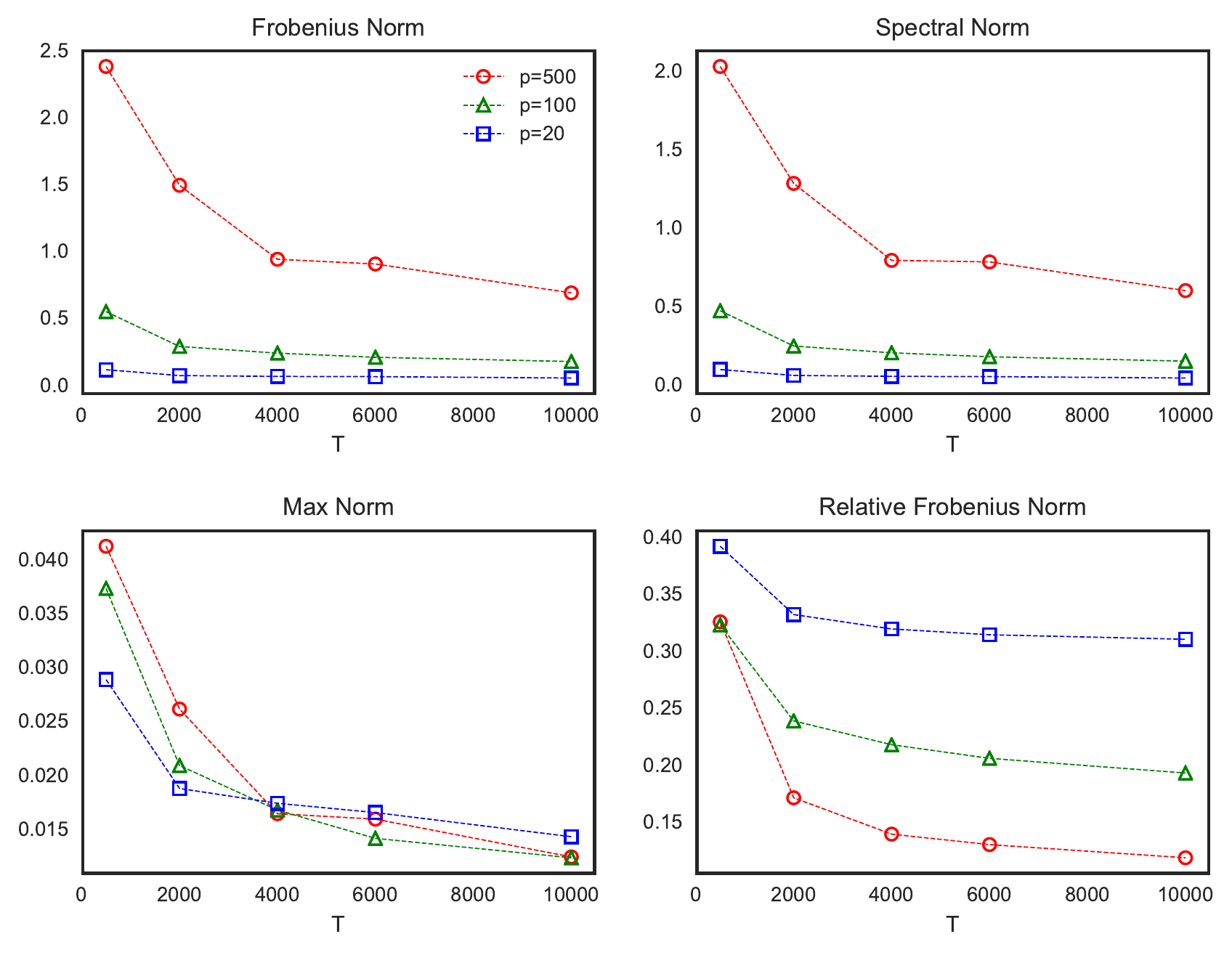}
\caption{Average predicted volatility errors of $\hat{\bSigma}_{t+1}-\bSigma_{t+1}$ against $T$ under the Frobenius, spectral, max, and relative Frobenius norms  with $p=20,100,500$.}
\label{fig:error}
\end{figure}

With the estimated parameter $\hat{\btheta}$, we validated the one-step ahead volatility prediction $\hat{\bSigma}_{t+1}$ with several matrix norms. 
We implemented the thresholding procedure for the idiosyncratic volatility in \eqref{eq:thresholding} with the tuning parameters $C_{\tau}=1$ and $s_p=1$.
Figure \ref{fig:error} displays the prediction errors of $\hat{\bSigma}_{t+1} - \bSigma_{t+1}$ with the Frobenius, spectral, max, and relative Frobenius norms.
We find that  the errors are large at $p=500$ for the Frobenius and spectral norms that depend on the dimensionality $p$, whereas the error of the relative Frobenius norm decreases as the dimension p increases. 
This result thus highlights that the relative Frobenius norm can lead to  the blessing of dimensionality, thereby helping risk managers enjoy the use of financial big data for   volatility prediction. 
The max norm shows no significant difference for different $p$, which can be explained by the fact that a large volatility matrix is more likely to have a large max norm error, despite a decrease in the element-wise error. 
 In all cases, the errors decrease as the sample size $T$ increases, which supports the theoretical findings in Theorem \ref{thm:prediction}.

\begin{figure}[h!]
\centering
\includegraphics[width = \textwidth]{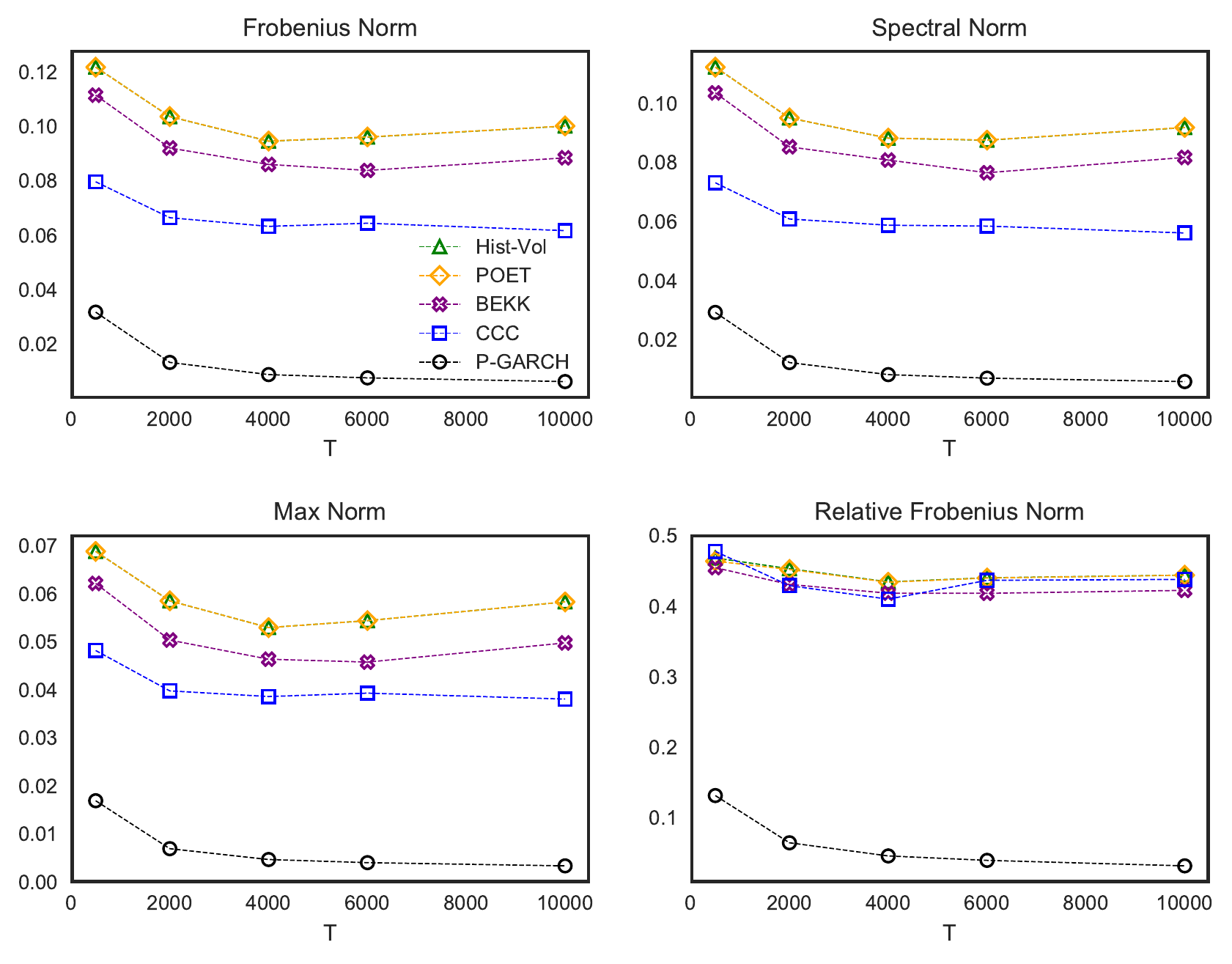}
\caption{Average predicted volatility errors of the portfolio volatility matrix $\hat{\bSigma}_{s,t+1}-\bSigma_{s,t+1}$  against $T$ under the Frobenius, spectral, max, and relative Frobenius norms with $p=500$ and portfolio size 5.}
\label{fig:error_compare}
\end{figure}

We compared the performance of predicting one-step ahead volatility matrices of the assets in the portfolio to clarify the benefits of using financial big data for small portfolio risk analysis
We set the portfolio size as $s=5$ and  the total number of assets as $p=500$. 
To capture the market dynamics within the portfolios of interest and compare with our proposed model (P-GARCH), we considered the CCC model \citep{bollerslev1990modelling}, and BEKK with diagonal-constraint  and variance targeting  \citep{engle1995multivariate,pedersen2014multivariate}.
Additionally, in the empirical study, we often impose the static or slow time varying covariance assumption, and under this condition, we adopted POET \citep{fan2013large} and historical sample covariance (Hist-Vol).
Figure  \ref{fig:error_compare} plots the average predicted volatility errors of the portfolio volatility matrix $\hat{\bSigma}_{s,t+1}-\bSigma_{s,t+1}$ against the sample size $T$ under the Frobenius, spectral, max, and relative Frobenius norms.
We observe that the  P-GARCH model is superior to the other multivariate volatility matrix models by showing much lower errors across different sample sizes.
This finding can be explained by the fact that the P-GARCH can address the market dynamics with small portfolios, whereas the others cannot fully address dynamics.


Finally, we compared the VaR forecasting performance with the small portfolio risk analysis methods.
We set $\alpha$ as 1\% and fixed $p=500$ and varied the portfolio size from 1 to 20.
We also calculated the small portfolio matrix estimators $\hat{\bSigma}_{s,t+1}$  based on the P-GARCH, CCC,   BEKK, POET, and Hist-Vol for the volatility forecast.
We compared the considered models with the portfolio univariate GARCH(1,1) model, which is a widely-used approach to modeling a portfolio return \citep{bollerslev1986generalized}. 
For the $\alpha$-quantile, we considered the normal and student-t distribution with 6 degrees of freedom ($\nu=6$) for the parametric method, whereas we used the non-parametric method with $\lceil \alpha T \rceil$-th smallest values of the standardized portfolio log returns, as in Section \ref{sect:forecast:VaR}.
The portfolio was set to be equally weighted.
The true VaR is $z_\alpha (\bw^\T \bSigma_{t+1}\bw)^{1/2}$, where $z_\alpha$ is the $\alpha$-quantile for the standard normal distribution.

\begin{figure}[h!]
\centering
\includegraphics[width = \textwidth]{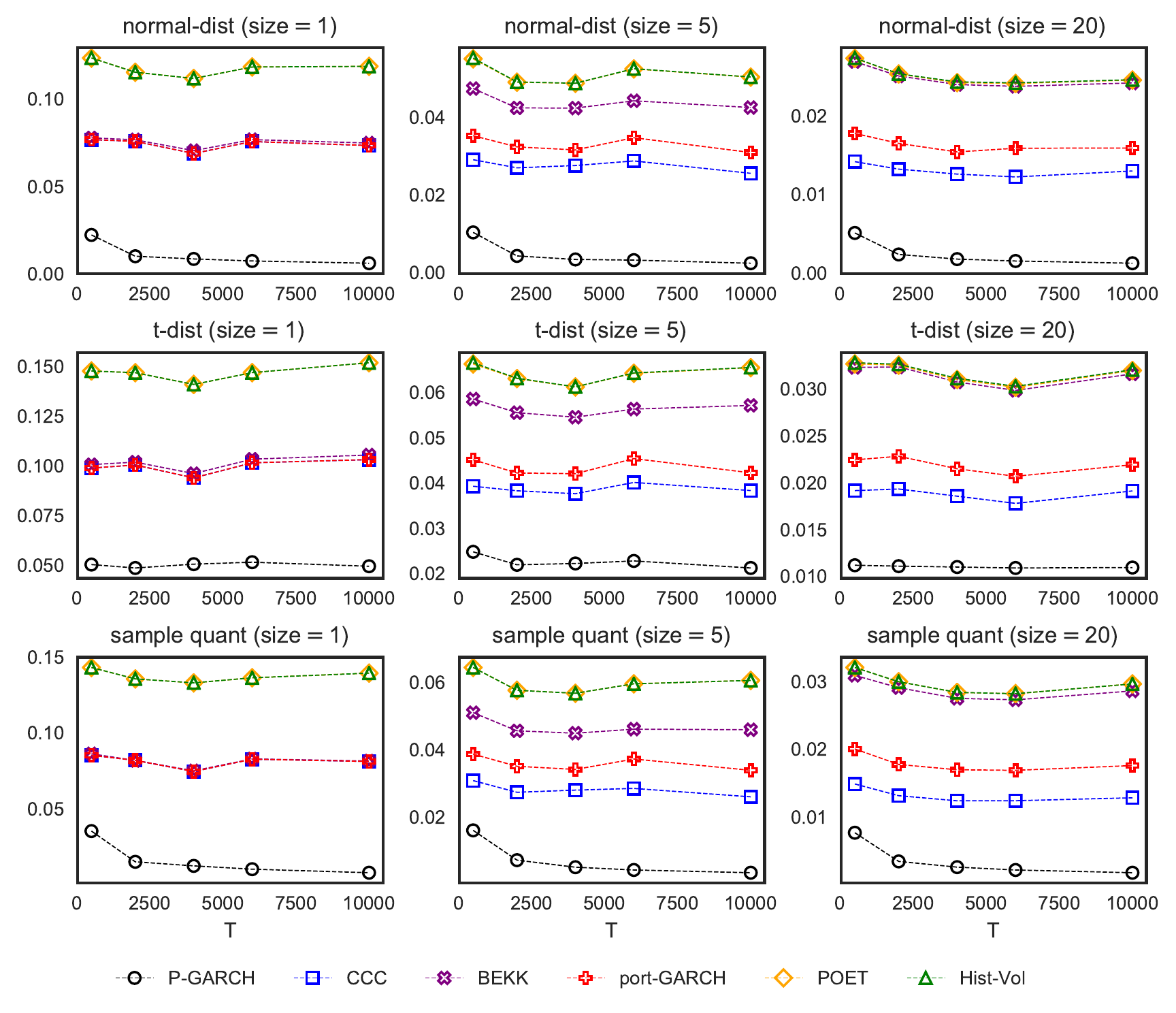}
\caption{MAE of the 1\%-level VaR forecasts for $p=500$ with 500 replications. From the left, the plot shows the MAE of VaR with the parametric methods under normal and t-distribution ($\nu=6$) and non-parametric method with the sample quantile with the portfolio sizes 1, 5, and 20.}
\label{fig:error_VaR}
\end{figure}

Figure  \ref{fig:error_VaR} shows the MAEs of the 1\%-level VaR forecasts for $p=500$ with 500 replications.
We identify that the P-GARCH outperforms the other models for any $\sigma$-based methods and portfolio size. 
When comparing the quantile estimation methods, given that the true model is based on the normal distribution, the normal distribution-based method shows the best performance.
However, the sample quantile-based method shows similar performance as the sample size $T$ increases. 
This is because the sample quantile is a non-parametric consistent estimation procedure; thus, with a large $T$,  we can estimate the quantile consistently. 
Finally, the parametric-based methods, such as  P-GARCH, CCC,   BEKK, and port-GARCH,  usually perform better than the non-parametric methods. 
However, as the portfolio size increases, the BEKK method performs worse. 
One possible explanation is that as the portfolio size increases, the BEKK model becomes too complex to estimate model parameters.

\subsection{Empirical study} 	\label{sect:empirics}

In this section, we examined the performance of one-step ahead VaR forecasting  with the empirical data.
The data comprise 18 years CRSP daily percentage log returns of S\&P 500 constituents from January 1, 2000, to December 31, 2017 (4,523 days). 
The log returns are calculated from the percentage return of the last sale or closing bid/ask price including dividends.
We selected firms that have been ever a constituent of the S\&P 500 during this period and filtered out the stocks that have any missing data.
This selection process led us to include 492 firms ($p=492$) for the empirical study. 

\begin{figure}[h!]
\centering
\includegraphics[width=1\textwidth]{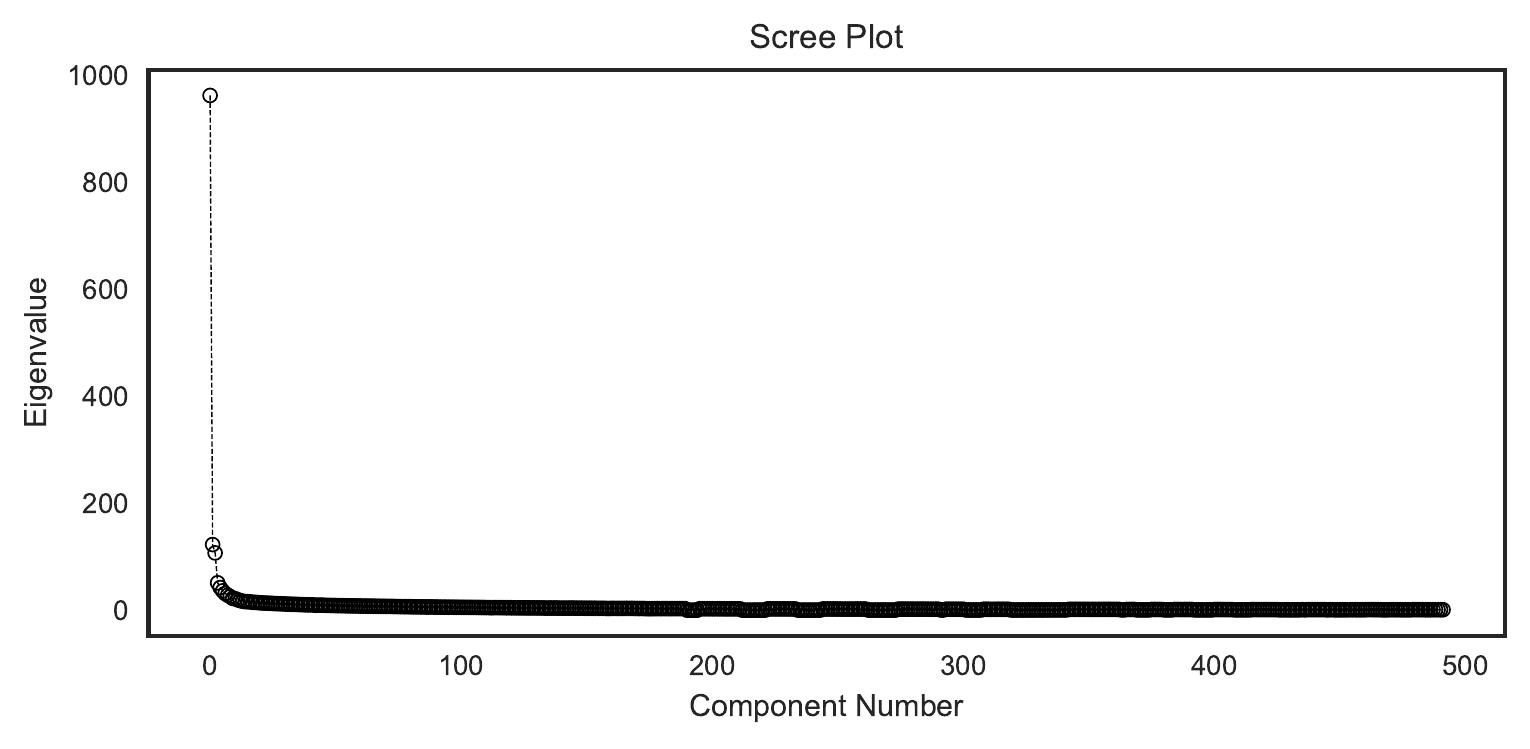}
\caption{\small Scree plot of the sample covariance matrix of 492 stocks.}
\label{fig:scree_plot}
\end{figure}

\begin{figure}[h]
\centering
\includegraphics[width=1\textwidth]{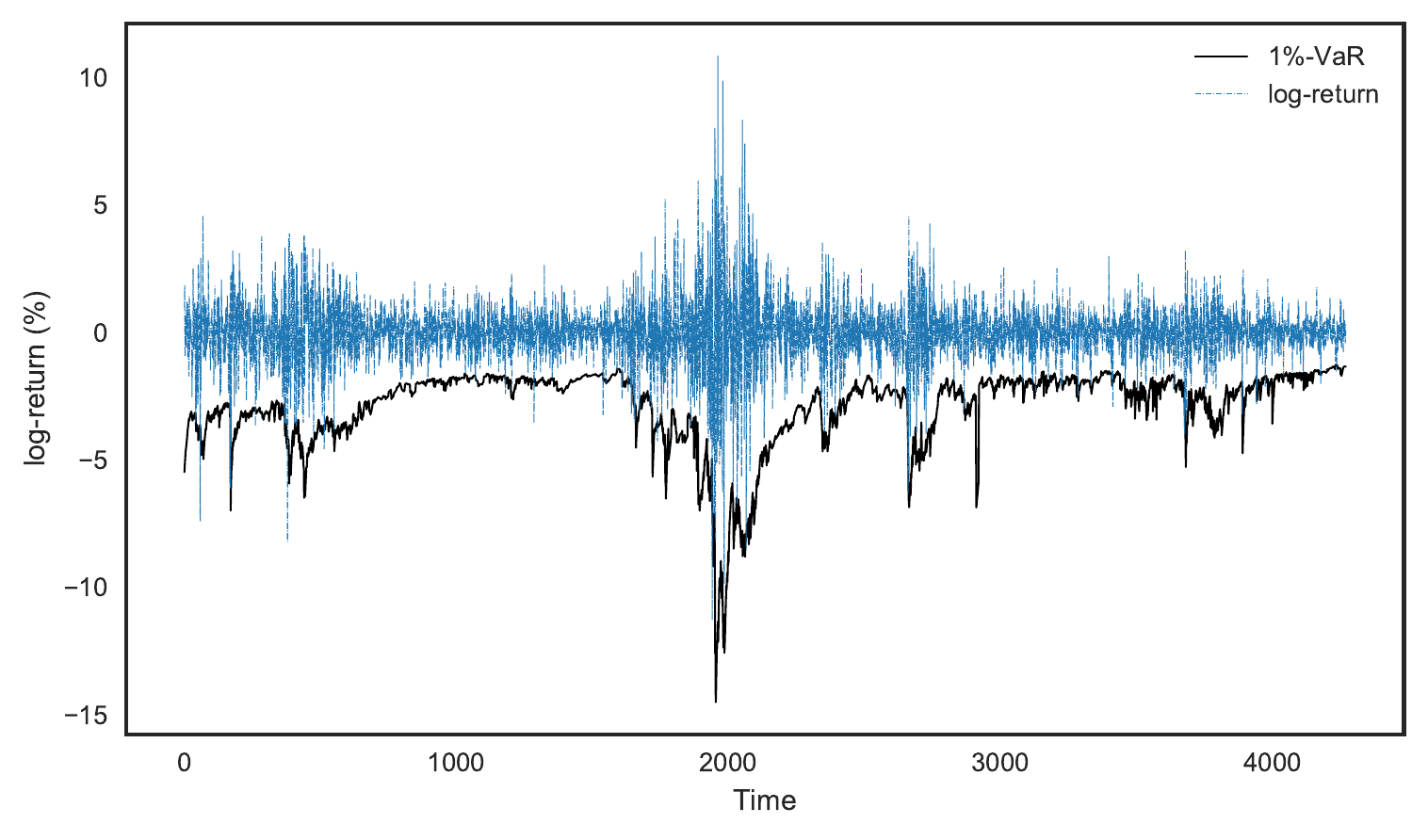}
\caption{\small Plots of daily log returns and predicted one-step ahead VaRs of the P-GARCH model. }
\label{fig:VaR_path}
\end{figure}

To employ the proposed P-GARCH model, we should determine the number of factors ($r$).
Figure 4 shows that the optimal $r$ is less than $3$,  and the rank $\hat{r}$  estimated  by \eqref{eq-rank-est} in the online supplement is also 3. 
Thus, we determined the possible values of $r$ as 1, 2, and 3, which is in line with  previous empirical studies \citep{chan1998risk,fama1992cross,fama1993common}. 
For the thresholding step, we used the global industry classification standard (GICS) sector \citep{fan2016incorporating}.
For example, we maintained within-sector volatilities but set others to zero.
The equally weighted portfolios of sizes 5 and 20 are randomly sampled from the 492 stocks with 500 repetitions, and the single-asset portfolio contains only one stock from 492 assets.
With these portfolios, we predicted VaR with a significance level of $\alpha=10\%$, 5\%, 2\%, and 1\% using the parametric and non-parametric methods, as discussed in Section \ref{sect:forecast:VaR}.
For the model fitting, we employed a rolling window scheme with the window size $T=252$ days (1 year).
The parameter is updated every 10 days, whereas the VaR forecasting is done every day with the updated log return data.
The total forecasting number $N$ is 4,270. 
The estimated parameters of P-GARCH(r=3) for the entire period are $\hat{\btheta}=(\hat{\bomega},\vec(\hat{\bA}),\vec(\hat{\bB}))=10^{-3}\times($16.32, 0.32, 0.42, 83.99, 0.00, 0.08, 6.65, 48.23, 0.90, 87.04, 5.88, 62.87, 892.7, 0.00, 0.00, 0.01, 944.7, 0.23, 0.01, 0.00, 934.99$)$.
The estimated parameter matrix $\hat{\bB}$ is almost diagonal, which indicates that the recursive structure mainly comes from its own volatility. 
Similarly, the estimated parameter matrix $\hat{\bA}$ is also almost diagonal, except for $A_{13}$.
However, when considering the magnitudes of the first and third eigenvalues, $A_{13}$ does not affect the first eigenvalue significantly.
Thus, the dynamics of the three factors are mainly  explained by their own past volatilities. 
Figure \ref{fig:VaR_path} depicts one sample path of the estimated one-step ahead 1\%-level VaR under t-distribution with a portfolio size of 5.
It shows that predicted VaR of the P-GARCH model can capture the dynamics of extreme loss.

\begin{figure}[h]
\centering
\includegraphics[width=1\textwidth]{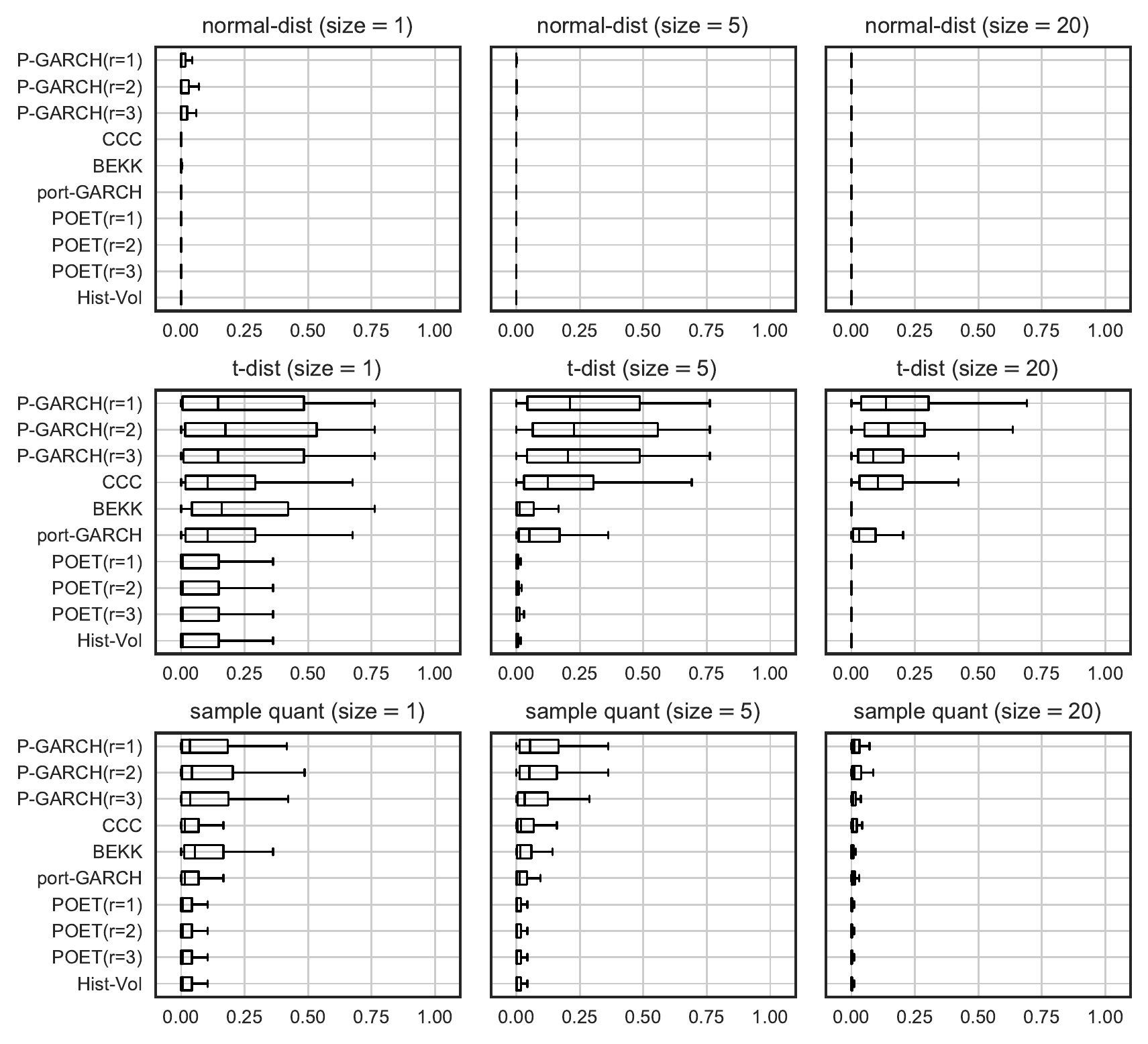}
\caption{\small Box plots of individual portfolio's $p$-values for $LR_{cc}$ with $\alpha=1\%$. 
Each column shows three different VaR estimation results with the normal, t-distribution ($\nu=6$), and sample quantile. 
Each row shows three different VaR estimates for the portfolio sizes 1, 5, 20.}
\label{fig:box_plot_size1}
\end{figure}

 We compared the out-of-sample VaR forecast performances with the small portfolio risk analysis methods considered in Section \ref{sect:Monte}. 
For example, we used the P-GARCH, CCC, BEKK, portfolio GARCH, Hist-Vol, and POET for the volatility prediction and the standard normal quantile, t-quantile with degrees of freedom 6, and sample quantile for the $\alpha$-quantile.
To test whether the predicted VaR is correct, we used the unconditional coverage test ($LR_{uc}$) \citep{kupiec1995techniques}, the conditional coverage test ($LR_{cc}$) \citep{christoffersen1998evaluating}, and the dynamic quantile test (DQ$_{\text{VaR}}$) \citep{kuester2006value}.   
Table \ref{table:VaR_1}  reports the averaged hit rates ($N_{\alpha}/N$) and the average $p$-values of the $LR_{uc}, LR_{cc}$, DQ$_{\text{Hit}}$, and DQ$_{\text{VaR}}$ for  portfolio sizes of 1, 5, 20 and $\alpha=1\%$ with the normal distribution, student-t distribution, and sample quantile.
 Figure \ref{fig:box_plot_size1} presents box plots of individual portfolio's LR$_{cc}$ $p$-values for  portfolio sizes of 1, 5, and 20 and $\alpha=1\%$ with the normal distribution, t-distribution, and sample quantile.
 From Table \ref{table:VaR_1} and  Figure \ref{fig:box_plot_size1},   we find that when comparing the dynamic models and static models, the dynamic models, such as the P-GARCH, BEKK, CCC, and portfolio GARCH, show better performance than the non-parametric models.
This indicates that the financial market is not static and that the GARCH-type models can account for the market dynamics.
When   the dynamic models are compared,  the P-GARCH has the highest $p$-values for most of the VaR tests.
From this empirical result, we can conjecture  that the P-GARCH accounts for the market dynamics by incorporating financial big data, and this leads to improved  performance in VaR forecasting for the relatively small portfolio.
Finally, when   the quantile methods are compared, the t-distribution shows the best performance.
Moreover,  the normal distribution has the smallest $p$-values for all the portfolio sizes.
This pattern coincides with the stylized fact that the log return follows the conditional heavy-tailed distribution \citep{cont2001empirical}.

\begin{landscape}
\begin{table}[h]
\begin{singlespace}
\caption{Average hit rate ($N_\alpha/N$) and average $p$-values of the $1\%$-level VaR test statistics under the normal, t-distribution $\nu=6$ and sample quantile for portfolio sizes of 1, 5, and 20.}
\vspace{2mm}
\label{table:VaR_1}
\end{singlespace}
\centering
\renewcommand\arraystretch{1}
\footnotesize
\setlength\tabcolsep{3pt}
\scalebox{1}{
\begin{tabular}{crrrrrrrrrrrrrrrrr}
\thickhline
\multirow{1}{*}{Models} & \multicolumn{1}{c}{$N_\alpha / N$} &  \multicolumn{1}{c}{$\text{LR}_{uc}$} & \multicolumn{1}{c}{$\text{LR}_{cc}$} &  \multicolumn{1}{c}{DQ$_{\text{Hit}}$}  &  \multicolumn{1}{c}{DQ$_{\text{VaR}}$}&& \multicolumn{1}{c}{$N_\alpha / N$} &  \multicolumn{1}{c}{$\text{LR}_{uc}$} & \multicolumn{1}{c}{$\text{LR}_{cc}$} &  \multicolumn{1}{c}{DQ$_{\text{Hit}}$}  &  \multicolumn{1}{c}{DQ$_{\text{VaR}}$}&& \multicolumn{1}{c}{$N_\alpha / N$} &  \multicolumn{1}{c}{$\text{LR}_{uc}$} & \multicolumn{1}{c}{$\text{LR}_{cc}$} &  \multicolumn{1}{c}{DQ$_{\text{Hit}}$}  &  \multicolumn{1}{c}{DQ$_{\text{VaR}}$}   \\
\hline
& \multicolumn{5}{c}{normal-distribution ($\text{size}=1$)} && \multicolumn{5}{c}{normal-distribution ($\text{size}=5$)} && \multicolumn{5}{c}{normal-distribution ($\text{size}=20$)} \\
\cline{2-6} \cline{8-12} \cline{14-18}
P-GARCH(r=1) &        0.016 &          0.043 &          0.040 &            0.024 &            0.021 &                    &        0.018 &          0.002 &          0.003 &            0.001 &            0.001 &                     &        0.019 &          0.000 &          0.000 &            0.000 &            0.000 \\
P-GARCH(r=2) &        0.015 &          0.049 &          0.047 &            0.032 &            0.029 &                    &        0.017 &          0.004 &          0.004 &            0.001 &            0.001 &                     &        0.019 &          0.000 &          0.000 &            0.000 &            0.000 \\
P-GARCH(r=3) &        0.015 &          0.045 &          0.042 &            0.028 &            0.025 &                    &        0.017 &          0.002 &          0.002 &            0.000 &            0.001 &                     &        0.019 &          0.000 &          0.000 &            0.000 &            0.000 \\
CCC          &        0.018 &          0.002 &          0.004 &            0.003 &            0.001 &                    &        0.018 &          0.000 &          0.000 &            0.000 &            0.000 &                     &        0.019 &          0.000 &          0.000 &            0.000 &            0.000 \\
BEKK         &        0.017 &          0.006 &          0.009 &            0.007 &            0.004 &                    &        0.020 &          0.000 &          0.000 &            0.000 &            0.000 &                     &        0.022 &          0.000 &          0.000 &            0.000 &            0.000 \\
port-GARCH   &        0.018 &          0.003 &          0.004 &            0.003 &            0.001 &                    &        0.019 &          0.000 &          0.000 &            0.000 &            0.000 &                     &        0.021 &          0.000 &          0.000 &            0.000 &            0.000 \\
POET(r=1)    &        0.018 &          0.008 &          0.007 &            0.001 &            0.001 &                    &        0.021 &          0.000 &          0.000 &            0.000 &            0.000 &                     &        0.023 &          0.000 &          0.000 &            0.000 &            0.000 \\
POET(r=2)    &        0.018 &          0.008 &          0.007 &            0.001 &            0.001 &                    &        0.020 &          0.000 &          0.000 &            0.000 &            0.000 &                     &        0.022 &          0.000 &          0.000 &            0.000 &            0.000 \\
POET(r=3)    &        0.018 &          0.008 &          0.007 &            0.001 &            0.001 &                    &        0.020 &          0.000 &          0.000 &            0.000 &            0.000 &                     &        0.022 &          0.000 &          0.000 &            0.000 &            0.000 \\
Hist-Vol     &        0.018 &          0.008 &          0.007 &            0.001 &            0.001 &                    &        0.020 &          0.000 &          0.000 &            0.000 &            0.000 &                     &        0.022 &          0.000 &          0.000 &            0.000 &            0.000 \\
[.1cm]
& \multicolumn{5}{c}{t-distribution ($\text{size}=1$)} && \multicolumn{5}{c}{t-distribution ($\text{size}=5$)} && \multicolumn{5}{c}{t-distribution ($\text{size}=20$)} \\
\cline{2-6} \cline{8-12} \cline{14-18}
P-GARCH(r=1) &        0.011 &          0.367 &          0.243 &            0.110 &            0.105 &                     &        0.012 &          0.339 &          0.279 &            0.039 &            0.044 &                    &        0.012 &          0.208 &          0.206 &            0.005 &            0.005 \\
P-GARCH(r=2) &        0.011 &          0.405 &          0.262 &            0.121 &            0.117 &                     &        0.012 &          0.377 &          0.309 &            0.060 &            0.061 &                    &        0.012 &          0.190 &          0.202 &            0.014 &            0.012 \\
P-GARCH(r=3) &        0.011 &          0.390 &          0.246 &            0.120 &            0.118 &                     &        0.012 &          0.353 &          0.272 &            0.053 &            0.053 &                    &        0.013 &          0.150 &          0.154 &            0.012 &            0.012 \\
CCC          &        0.013 &          0.202 &          0.195 &            0.093 &            0.060 &                     &        0.013 &          0.212 &          0.208 &            0.012 &            0.016 &                    &        0.013 &          0.125 &          0.153 &            0.001 &            0.001 \\
BEKK         &        0.012 &          0.264 &          0.250 &            0.090 &            0.078 &                     &        0.014 &          0.065 &          0.066 &            0.000 &            0.001 &                    &        0.016 &          0.004 &          0.004 &            0.000 &            0.000 \\
port-GARCH   &        0.013 &          0.202 &          0.195 &            0.093 &            0.061 &                     &        0.013 &          0.114 &          0.128 &            0.012 &            0.008 &                    &        0.014 &          0.048 &          0.072 &            0.001 &            0.001 \\
POET(r=1)    &        0.013 &          0.201 &          0.107 &            0.027 &            0.021 &                     &        0.015 &          0.031 &          0.023 &            0.000 &            0.000 &                    &        0.016 &          0.003 &          0.002 &            0.000 &            0.000 \\
POET(r=2)    &        0.013 &          0.201 &          0.107 &            0.027 &            0.021 &                     &        0.015 &          0.036 &          0.027 &            0.000 &            0.000 &                    &        0.016 &          0.004 &          0.003 &            0.000 &            0.000 \\
POET(r=3)    &        0.013 &          0.201 &          0.107 &            0.027 &            0.021 &                     &        0.015 &          0.036 &          0.028 &            0.000 &            0.000 &                    &        0.016 &          0.004 &          0.003 &            0.000 &            0.000 \\
Hist-Vol     &        0.013 &          0.201 &          0.107 &            0.027 &            0.021 &                     &        0.015 &          0.032 &          0.024 &            0.000 &            0.000 &                    &        0.016 &          0.003 &          0.002 &            0.000 &            0.000 \\
[.1cm]
& \multicolumn{5}{c}{sample quantile ($\text{size}=1$)} && \multicolumn{5}{c}{sample quantile ($\text{size}=5$)} && \multicolumn{5}{c}{sample quantile ($\text{size}=20$)} \\
\cline{2-6} \cline{8-12} \cline{14-18}
P-GARCH(r=1) &        0.013 &          0.172 &          0.130 &            0.050 &            0.032 &                   &        0.013 &          0.104 &          0.120 &            0.029 &            0.028 &                     &        0.014 &          0.024 &          0.031 &            0.004 &            0.005 \\
P-GARCH(r=2) &        0.013 &          0.183 &          0.144 &            0.066 &            0.041 &                   &        0.013 &          0.097 &          0.112 &            0.033 &            0.031 &                     &        0.014 &          0.021 &          0.034 &            0.009 &            0.007 \\
P-GARCH(r=3) &        0.013 &          0.152 &          0.125 &            0.050 &            0.031 &                   &        0.013 &          0.075 &          0.088 &            0.025 &            0.027 &                     &        0.015 &          0.012 &          0.021 &            0.005 &            0.005 \\
CCC          &        0.014 &          0.042 &          0.059 &            0.031 &            0.013 &                   &        0.014 &          0.043 &          0.058 &            0.003 &            0.004 &                     &        0.015 &          0.014 &          0.023 &            0.000 &            0.000 \\
BEKK         &        0.013 &          0.107 &          0.125 &            0.047 &            0.027 &                   &        0.014 &          0.054 &          0.063 &            0.001 &            0.001 &                     &        0.014 &          0.017 &          0.010 &            0.000 &            0.000 \\
port-GARCH   &        0.014 &          0.042 &          0.059 &            0.031 &            0.013 &                   &        0.014 &          0.024 &          0.039 &            0.005 &            0.003 &                     &        0.015 &          0.007 &          0.012 &            0.000 &            0.000 \\
POET(r=1)    &        0.013 &          0.107 &          0.071 &            0.015 &            0.006 &                   &        0.014 &          0.043 &          0.029 &            0.000 &            0.000 &                     &        0.015 &          0.016 &          0.008 &            0.000 &            0.000 \\
POET(r=2)    &        0.013 &          0.107 &          0.071 &            0.015 &            0.006 &                   &        0.014 &          0.043 &          0.029 &            0.000 &            0.000 &                     &        0.015 &          0.016 &          0.008 &            0.000 &            0.000 \\
POET(r=3)    &        0.013 &          0.107 &          0.071 &            0.015 &            0.006 &                   &        0.014 &          0.043 &          0.029 &            0.000 &            0.000 &                     &        0.015 &          0.016 &          0.008 &            0.000 &            0.000 \\
Hist-Vol     &        0.013 &          0.107 &          0.071 &            0.015 &            0.006 &                   &        0.014 &          0.043 &          0.029 &            0.000 &            0.000 &                     &        0.015 &          0.016 &          0.008 &            0.000 &            0.000 \\
\hline
\end{tabular}
}
\end{table}
\end{landscape}

\section{Conclusion} 	\label{sect:conclusion}

In this paper, we aim to develop  a dynamic process of portfolio risk measurement with the use of financial big data. 
The proposed model benefits from the use of financial big data in that it leads risk managers to take into account latent risk factors that are not included in their portfolios. 
One may be concerned about the curse of dimensionality with regard to the use of financial big data in parameter estimation; however, the proposed approach can overcome this problem with the POET method \citep{fan2013large} under the dynamic factor model in a GARCH form. 
We showed  that a large number of assets aids in increasing the accuracy of factor volatility estimation.
This finding results from the fact that every asset contains common market information so that an increase in the number of assets enables one to obtain more information on latent risk factors. 
Thus, the proposed model achieves the blessing of dimensionality in estimating latent common factors.

We tested a finite sample performance of the proposed dynamic volatility model and the corresponding VaR model with Monte Carlo simulations and showed that the results support our theoretical findings.
 We further investigated the performance of our one-step ahead volatility matrices in the asset portfolio and concluded that our dynamic process with P-GARCH outperforms other competitive models for any $\sigma$-based methods and portfolio size. 
 To guarantee the performance of our predictive model with financial big data, we used S\&P 500 constituents with  parametric and non-parametric methods. 
 The results illustrated that the predicted VaR of our dynamic volatility model can capture the dynamics of extreme loss, which can be extreme cases from systematic factors in the market.
  Finally, the empirical study shows that our dynamic model with P-GARCH outperforms the other competitive dynamic models in terms of forecasting the VaR with financial big data for a relatively small portfolio.
 
The lesson from the 2008 financial crisis and similar systemic events in the financial market of history clearly argues that risk managers in the financial industry may have to take into account latent risks possibly accounting for systematic factors. 
The portfolios that they aim to analyze may consist of idiosyncratic factors with a limited number of risk factors (or assets); hence, the predicted level of risk in their target portfolios may fail to address the entire risk landscape. 
This problem can be also of importance for financial regulators in that underestimation of risk capitals based on a small portfolio risk measurement would potentially prevent the financial markets from being stabilized, thereby leading to negative externality to the economy, as observed in 2008 \citep{kaserer2019systemic}\footnote{The negative externality was realized in 2008 as a huge bailout to, e.g., AIG as described in the introduction. Most of the \$374 billion total bailout was provided to financial firms, e.g., AIG, Citigroup and Bank of America, and the government rationale of this financial support was to avoid the collapse of world financial markets potentially creating the risk of another Great Depression \citep{harrington2009financial}. This negative externality to the economy clearly burdened tax-payers for not only the size of bailout from taxes, but also the negative spillover effect on real economy.}.  
This study contributes to this regard by providing an innovative approach to using financial big data in improve the accuracy of portfolio risks.

In this paper, we only consider the factor dynamics, but   several studies have also shown that the idiosyncratic volatilities are also affected by common market factors  \citep{barigozzi2016generalized, connor2006common, herskovic2016common, rangel2012factor, shin2021factor}. 
Thus, it is interesting but difficult to develop idiosyncratic dynamic models. 
In addition,  we consider the $\sigma$-based approach to estimate the VaR values with the parametric and non-parametric quantile structures.
However, the parametric structure cannot account for some stylized features, such as  the skewness of the asset returns, whereas  the empirical quantile approach often has large errors in the extreme tail. 
Thus, it is important and interesting to develop a quantile estimation procedure that can accomodate the stylized features of the asset returns.
Finally, we assume that the mean process is constant over time, but this assumption is often violated in finance practice. 
Unlike the volatility process, the mean process does not have a strong linear time series structure, which makes the modeling of a dynamic mean process difficult. 
Thus,  a parametric structure of the mean process, which can account for the heterogeneity of the mean process, should be developed.
We leave these interesting problems for future study.

\section*{Online supplement}
Theorems and technical proofs are provided online at Wiley in  the Supplementary Material of this article.
Readers may refer to the supplementary material associated with this article, available at Wiley (\url{https://onlinelibrary.wiley.com/journal/15396975}).


\clearpage

\let\normalsize\small
\appendix
\small

\vspace*{0.3cm}
\begin{center}
\linespread{1.6}\Large  Supplementary Material on “Next Generation Models for Portfolio Risk Management: An Approach Using Financial Big Data”
\end{center}
\vspace{1cm}


This appendix provides micelleneous materials, theorems, and technical proofs of the paper entitled “Next Generation of Portfolio Risk Management: An Approach to Using Financial Big Data”.

We first introduce some notations. 
For a symmetric matrix $\bA$,   $\lambda_k(\bA)$ is the $k^{th}$ largest eigenvalue of $\bA$, $\lambda_{\max}(\bA)$ is the largest eigenvalue, and $\lambda_{\min}(\bA)$ is the smallest eigenvalue. 
$\| \cdot\|_\infty$ is the matrix $\ell_\infty$ norm.
Denote $ \bA\circ\bB$ the Hadamard (or element-wise) product and $\bx^{2}=\bx \circ \bx$ the element-wise square of a vector $\bx$.
$\tr(\bA)$ and   $\text{vec}(\bA)$ are the trace and vectorization of $\bA$, respectively. 
Denote $\ind_{E}$ the indicator function of the event $E$ and $\be_i$ the $i^{th}$ standard basis vector. 
To simplify the notations, we define $\partial$ as a partial derivatives: $\partial_x y(\bx) = \partial y(\bx)/\partial{x}$; $\partial_x \by(\bx) = \partial \by(\bx)/\partial x=[\partial y_1/\partial x,\ldots, \partial y_{\dim(\by)}/\partial x]^\T$;  $\partial_{\bx} y(\bx) =[\partial y/\partial x_1,\ldots, \partial y/\partial x_{\dim(\bx)}]$; $\partial_{\bx} \by(\bx) = [\partial \by/\partial x_1,\ldots, \partial \by/\partial x_{\dim(\bx)}]$, where the functions $y(\bx)$ and $\by(\bx)$ are differentiable. 
$C$ is a generic constant that may differ from each equations.

\renewcommand{\thesection}{S\arabic{section}}  
\renewcommand{\thetable}{S\arabic{table}}  
\renewcommand{\theequation}{S\arabic{equation}}
\renewcommand{\thefigure}{S\arabic{figure}}

\section{Choice of the rank}

 In this paper, we assume that the rank $r$ is known. 
 However, in practice, the rank is unknown, so it is important to estimate the rank.
\citet{ait2017using} introduced the rank estimation procedure as follows:
 \begin{equation}\label{eq-rank-est}
	\hat{r} = \arg \min_{1 \leq j \leq r_{\max} }   p^{-1} \hat{\lambda}_{ j} + j \times c_1  \left \{ \sqrt{ \log p / T }  + p^{-1} \log p \right \} ^{c_2}  -1,
\end{equation}
where $ r_{\max}$, $ c_1$, and $c_2$  are  tuning parameters. 
Under some technical conditions, we can show its consistency. 
We note that the overestimated $\hat{r}$ is not harmful because the factor model includes the important factors and the impact of the misspecified  factors is relatively small \citep{fan2013large}. 
In contrast, when the rank is underestimated, the factor model losses important information.

\section{Theorems for the latent factor estimation} \label{proof:thm:factor}

\begin{assumption}    ~
\label{ass:factor}
\begin{enumerate}
\item [(a)] $\bbE\left[ f_{it}^8 \right] ,  \bbE\left[h_{it}^4(\btheta_0) \right] , \bbE\left[ \left( u_{it} u_{jt}- \Sigma_{u,ij} \right)^4 \right]$, and $\bbE\big[\big( \bq^\T \bu_t / \sqrt{\bq^\T \bSigma_u \bq} \big)^4 \big]$ are bounded by some constant $C$ for all $i,j$ and $\bq$ s.t. $\|\bq\|=1$.
\item [(b)] The minimum eigen-gap $\delta_{r}= \min_{k \le r}\big| \lambda_k(\bar{\bSigma}_f) - \lambda_{k+1}(\bar{\bSigma}_f) \big|$ satisfies $\delta_{r}  \ge C$ $\text{a.s.}$ 
\end{enumerate}
\end{assumption}

\begin{remark}
Assumption \ref{ass:factor}(b) is the pervasive condition that is widely used in the low-rank matrix inferences \citep{ait2017using,fan2013large,fan2018large,fan2019robust,kim2019factor,stock2002forecasting}.  
The pervasive condition with the sparsity condition \eqref{eq:sparsity} helps to distinguish the latent factor from the idiosyncratic volatility. 
Additionally, since the common market factor affects the whole asset, its proportion to the total variation is significant. 
Mathematically, this implies that the corresponding eigenvalues have the $p$ order, so the pervasive condition is not restrictive.   
\end{remark}

The following theorem provides the convergence rates of the latent factor component estimators $ \hat{\bV} $ and   $\hat{\bff}_t$.
\begin{thm}     \label{thm:factor}
Under the model in Section \ref{sect:model:factor},  suppose  Assumption \ref{ass:factor}. 
Then we have
\begin{align}
    &\min_{\bO}{\big\| \hat{\bV}  - \bV \bO \big\|^2}   =  O_p\left(\frac{p}{T} + \frac{s_p^2}{p} \right),     \label{eq:thm:factor:V}\\
    &\big\| \hat{\bff}_t^2 - \bff_t^2 \big\|   =  O_p\left( \frac{1}{\sqrt{T}} + \sqrt{\frac{s_p}{p}} \right),     \label{eq:thm:factor:f}
\end{align}
where $\bO$ is a diagonal sign matrix which has value $\pm 1$.
\end{thm}

\begin{remark}
Eigenvectors have the sign problem; for example, $-\bv_i$ and $\bv_i$ are not distinguishable, where $\bv_i$ is the $i^{th}$ column vector of $\bV$. So we put the sign matrix $\bO$ in  \eqref{eq:thm:factor:V} to identify the eigenmatrix. 
\end{remark}

\begin{remark}
Theorem \ref{thm:factor} shows that  the convergence rate of  $\hat{\bff}_t^2$ is $1/\sqrt{T}  + \sqrt{s_p/p}$. 
The first term $1/\sqrt{T}$ is the usual optimal convergence rate for estimating the mean conditional volatility matrix $\bar{\bSigma}$.
The second term $\sqrt{s_p/p}$ is the cost to identify the latent factor. 
This results match with the convergence rate derived in \citet{fan2011high, fan2013large, fan2018large}. 
We note that unlike the usual i.i.d. assumption, to manage the heterogeneous volatility structure, we impose the dynamic structure on the covariance matrix.  
Theorem \ref{thm:factor} shows the asymptotic properties under the heterogeneous structure.
\end{remark}

 \subsection{Proof of Theorem \ref{thm:factor}}

\begin{lemma} 	\label{lemma:avgyvol_error} 
Under the assumptions of Theorem \ref{thm:factor}, we have
\begin{align*}  
	\bbE\left\|\hat{\bar{\bSigma}}-\bar{\bSigma} \right\|_F^4 = O\left(\frac{p^4}{T^2} \right).
\end{align*}
\end{lemma}

\textbf{ Proof of Lemma \ref{lemma:avgyvol_error}.}
We have
\begin{align*}
	\hat{\bar{\bSigma}}-\bar{\bSigma} &= \frac{1}{T} \sum_{t=1}^T{\left((\by_t - \bar{\by})(\by_t - \bar{\by})^\T - \bSigma_t\right)}  	\\
	&= \frac{1}{T}\sum_{t=1}^T{\Big(\bV\left(\bff_t\bff_t^\T - \diag(\bh_{0,t})\right)\bV^\T + \bV \bff_t \bu_t^\T + \bu_t\bff_t^\T \bV^\T + \bu_t \bu_t^\T - \bSigma_u\Big)} 	\\
	&\quad\; - (\bar{\by} - \bmu)(\bar{\by} - \bmu)^\T,
\end{align*}
where $\bar{\by} = (1/T)\sum_{t=1}^T{\by_t}$ and $\bh_{0,t} = \bh_t (\btheta_0)$.
Then each element is a martingale difference.
Consider $\frac{1}{T} \sum_{t=1}^T  \bV\left(\bff_t\bff_t^\T - \diag(\bh_{0,t})\right) \bV^\T$.
We have
\begin{eqnarray*}
	  \bbE  \[  \(\frac{1}{T} \sum_{t=1}^T  ( f_{it} ^2  -  h_{0,it} ) \) ^4  \]   &\leq&  C \bbE  \[  \(\frac{1}{T} \sum_{t=1}^T  ( f_{it} ^2  -  h_{0,it} ) ^2 \) ^2  \]  \cr
	  &\leq& C T^{-2} ,
\end{eqnarray*}
where $ h_{0,it} $ is the $i$th element of  $\bh_{0,t}$, and the first inequality is due to the Burkholder-Davis-Gundy inequality \citep{chow2012probability}.
Thus, we have
\begin{eqnarray*}
	  \bbE  \[  \left \| \frac{1}{T}\sum_{t=1}^T  \bV\left(\bff_t\bff_t^\T - \diag(\bh_{0,t})\right)\bV^\T \right \|_F ^4  \]   &\leq& C p^4  \bbE  \[  \left \|  \frac{1}{T}\sum_{t=1}^T   \left(\bff_t\bff_t^\T - \diag(\bh_{0,t})\right)\  \right \|_F ^4  \]\cr  
	  &\leq& C p^{4} T^{-2}.
\end{eqnarray*}
We can show other elements similarly.
\endpf

\textbf{Proof of Theorem \ref{thm:factor}}.
First consider \eqref{eq:thm:factor:V}.
By the Davis-Kahan's sine theorem \citep{yu2015useful}, we have
\begin{align} 	\label{eq:V-V^2}
	\bbE\left\| \hat{\bV} - \bV\bO \right\|^4  &= p^2  \bbE\left\| \frac{1}{\sqrt{p}}\hat{\bV} - \frac{1}{\sqrt{p}}\bV\bO \right\|^4 	\nonumber \\
	&\le   p^2  \bbE\left[\frac{\left\| \hat{\bar{\bSigma}} - \bar{\bSigma} + \bSigma_u \right\|^4}{ p^4 \delta_{r}^4} \right] 	\nonumber \\
	&\leq C \( \frac{p^2}{T^2} + \frac{s_p^4}{p^2}\right),
\end{align}
where the last equality is due to Lemma \ref{lemma:avgyvol_error}  and the fact that
\begin{eqnarray*}
\|\bSigma_u\|  \le  \|\bSigma_u\|_\infty &\le&  \max_{j} \sum_{i=1}^p |\Sigma_{u, ij} | \cr
	&=& \max_{j} \sum_{i=1}^p |\Sigma_{u, ij} | ^q  |\Sigma_{u, ij} | ^{1-q}  \cr
	&\leq&	 C \max_{j} \sum_{i=1}^p |\Sigma_{u, ij} | ^q  \leq C s_p,
\end{eqnarray*} 
where the last inequality is due to the sparsity condition \ref{eq:sparsity}.

Consider \eqref{eq:thm:factor:f}. 
Without of loss of generality, we assume that $\bO$ is the identity matrix.
Algebraic manipulations show
\begin{align*}
	&\bbE\left\| \hat{\bff}_t^{2} - \bff_t^{2} \right\|^2 	\\
	&= \bbE\left\| \left(\frac{1}{p}\hat{\bV}^\T(\by_t - \bar{\by})\right)^{2} - \left(\frac{1}{p}\bV^\T(\by_t-\bmu-\bu_t)\right)^{2}  \right\|^2 	\\
	&= \frac{1}{p^4}\bbE \left\| (\hat{\bV}^\T \by_t)^2 - 2(\hat{\bV}^\T\by_t \circ \hat{\bV}^\T \bar{\by}) + (\hat{\bV}^\T \bar{\by})^2 - (\bV^\T\by_t)^2 + 2p \bff_t \circ \bV^\T(\bmu + \bu_t) + (\bV^\T(\bmu + \bu_t))^2 \right\| ^2	\\ 	
	&= \frac{1}{p^4}\bbE\Big\| (\hat{\bV} - \bV)^\T\by_t \circ (\hat{\bV}+\bV)^\T\by_t + 2p \bff_t \circ \bV^\T\bu_t  + (\bV^\T \bu_t)^{2} 	\\
	&\qquad\qquad - 2(\hat{\bV}^\T\by_t \circ \hat{\bV}^\T \bar{\by}) + (\hat{\bV}^\T \bar{\by})^2 + 2p \bff_t \circ \bV^\T\bmu + (\bV^\T \bmu)^2 + 2( \bV^\T \bu_t \circ \bV^\T \bmu) \Big\|^2 	\\
	&\le \frac{C}{p^4}\bigg(\bbE\left\| (\hat{\bV} - \bV)^\T\by_t \circ (\hat{\bV}+\bV)^\T\by_t\right\|^2 + p^2\bbE\left\|\bff_t \circ \bV^\T\bu_t \right\|^2  + \bbE\left\|(\bV^\T \bu_t)^{2} \right\|^2 	\\
	&\qquad\qquad + \bbE\Big\| - 2(\hat{\bV}^\T\by_t \circ \hat{\bV}^\T \bar{\by}) + (\hat{\bV}^\T \bar{\by})^2 + 2 (\bV^\T \by_t \circ \bV^\T \bmu) - (\bV^\T \bmu)^2 \Big\|^2 \bigg)	\\
	&= \frac{C}{p^4}\Big(\text{(I)} + \text{(II)} + \text{(III)} + \text{(IV)} \Big).
\end{align*}
For $\text{(I)}$, we have
\begin{align*}
	\text{(I)}  &= 	\bbE\left\| (\hat{\bV} - \bV )^\T\by_t \circ (\hat{\bV}+\bV )^\T\by_t\right\|^2\\
	&\le  \bbE\left[\left\|(\hat{\bV} - \bV)^\T \by_t \right\|^2 \left\|(\hat{\bV} + \bV)^\T \by_t \right\|^2 \right] 	\\
	&\le  \bbE\left[\left\|\hat{\bV} - \bV  \right\|^2 \left\|\hat{\bV} + \bV\right\|^2 \|\by_t\|^4\right] 	\\
	&\le  4p \,\bbE\left[\left\|\hat{\bV} - \bV \right\|^2  \|\by_t\|^4\right] 	\\
	&\le  4p \,\sqrt{\bbE\left\|\hat{\bV} - \bV \right\|^4} \sqrt{\bbE\|\by_t\|^8} 	\\
	&\le  C\left(\frac{p^4}{T} + p^2 s_p^2\right), 
\end{align*}
where  the fourth inequality is due to the H\"older's inequality and  the last inequality is due to  \eqref{eq:V-V^2}.
For $\text{(II)}$, we have
\begin{align*}
	\text{(II)}  &\le \bbE \|\bff_t\|^2 \bbE\left\| \bV^\T \bu_t \right\|^2  	\\
	&\le  C\left(p s_p \right),
\end{align*}
where the last inequality is by the fact that $\bbE\|\bV^\T\bu_t\|^2= \tr(\bV^\T \bSigma_u \bV) = O(p\,s_p)$.
For $\text{(III)}$, we have
\begin{align*}
	\text{(III)}  &=   \bbE\[ \sum_{i=1}^r (\bv_i^\T \bu_t)^{4}  \]   	\\
&= \bbE \left[ \sum_{i=1}^r \frac{(\bv_i^\T \bu_t)^4}{(\bv_i^\T \bSigma_u \bv_i)^2} \(\bv_i^\T \bSigma_u \bv_i  \)^2 \right] 	\\
	&\le C p^2 \| \bSigma_u \|^2  \bbE \left[ \sum_{i=1}^r \frac{(\bv_i^\T \bu_t)^4}{(\bv_i^\T \bSigma_u \bv_i)^2}   \right]  \\
	&\le C s_p^2 p^2,
\end{align*}
where $\bv_i$ is the $i^{th}$  column of $\bV$. 
 Similarly, we can show
\begin{align*}
	\text{(IV)}  &\le  2\bbE\Big\|-\hat{\bV}^\T \by_t \circ \hat{\bV} \bar{\by} + \bV^\T\bmu\circ \bV^\T\by_t \Big\|^2  + \bbE\Big\|(\hat{\bV}^\T \bar{\by})^2  - (\bV^\T \bmu)^2 \Big\|^2 	\\
	&\le C\left( \frac{p^4}{T} + p^2 s_p^2 \right) .
\end{align*}
Therefore, we have
$$\bbE\big\| \hat{\bff}_t^{2} - \bff_t^{2} \big\|^2 = O\left(1/T + s_p^2/p^2 + s_p/p \right).$$ \endpf

\section{Theorems for the QMLE estimation procedure} 	\label{proof:thm:GARCH}

Define 
\begin{align*}
	L_{0,T}(\btheta) &= -\frac{1}{T}\sum_{t=1}^T{\sum_{i=1}^r{\left(\log{h_{it}(\btheta)}  +  h_{it}^{-1}(\btheta)h_{0,it}\right)}} = \frac{1}{T}\sum_{t=1}^T{l_{0,t}(\btheta)}, 	\\
	L_{T}(\btheta) &= -\frac{1}{T}\sum_{t=1}^T{\sum_{i=1}^r{\left(\log{h_{it}(\btheta)}  +  h_{it}^{-1}(\btheta)f_{it}^2\right)}} = \frac{1}{T}\sum_{t=1}^T{l_{t}(\btheta)}, 	\\
	\hat{L}_{T}(\btheta) &= -\frac{1}{T}\sum_{t=1}^T{\sum_{i=1}^r{\left(\log{\hat{h}_{it}(\btheta)}  +  \hat{h}_{it}^{-1}(\btheta)\hat{f}_{it}^2\right)}} = \frac{1}{T}\sum_{t=1}^T{\hat{l}_{t}(\btheta)},
\end{align*}
where $\bh_{0,t} = \bh_t(\btheta_0)$ and $\hat{\bh}_t(\btheta) = \bomega + \bA \hat{\bff}_{t-1}^{2} + \bB \hat{\bh}_{t-1}(\btheta)$. 

\begin{assumption} ~  \label{ass:GARCH} 
\begin{enumerate}
\item[(a)]  The parameter space $\bTheta$ is a compact set such that every element $\btheta \in \bTheta$ is positive; $\sup_{\btheta \in \bTheta} \bbE[h_{it}^4(\btheta)]$ is bounded for all $i= 1,\ldots, r$ and $t =1,\ldots, T$; the eigenvalues of $\bB$ and $\bA$ are positive and $\|\bB\| <1$; and $\btheta_0$ is the interior point. 

\item[(b)] $\bff_t^2$'s are non-degenerate random variables.
\end{enumerate}
\end{assumption}

The below theorem illustrates the convergence rate of the QMLE estimator $\hat{\btheta}$. 
\begin{thm}     \label{thm:GARCH}
Under the model in Section \ref{sect:model:factor},  suppose  Assumptions \ref{ass:factor} and \ref{ass:GARCH} in the online supplement. 
Then we have
\begin{align}     \label{eq:GARCH_conv}
    \big\|  \hat{\btheta} - \btheta_0 \big\|_{\max} = O_p \left(\frac{1}{\sqrt{T}} + \sqrt{\frac{s_p}{p}}\right).
\end{align}
\end{thm}

\begin{remark} 
Theorem \ref{thm:GARCH} shows that the QMLE estimator $\hat{\btheta}$ has the convergence rate $1/\sqrt{T} + \sqrt{s_p/p}$. 
The term $\sqrt{s_p/p}$ originates from the latent factors estimation in \eqref{f-est}, which is the cost to identify the latent factor. 
Thus, when the factors are observable, the convergence rate will be $1/\sqrt{T}$.
\end{remark}

\subsection{Proof of Theorem \ref{thm:GARCH}}  

\begin{lemma}  	\label{lemma:irr}
Under the assumptions of Theorem \ref{thm:GARCH}, let $\bh_t$ to be a function of $\btheta$ and $\bh_1$,
$$
	\bh_t(\btheta, \bh_1) = \bB^{t-1} \bh_1 + (\bI_r - \bB)^{-1} \left(\bI_r - \bB^{t-1}\right)\bomega + \sum_{k=0}^{t-2}{\bB^k \bA \bff_{t-1-k}^{2}}.
$$
 Then, for some true initial value $\bh_{0,1}$, we have  
\begin{align*}
	\left\| \bh_t(\btheta, \bh_1) - \bh_t(\btheta, \bh_{0,1} )\right\|  &=  O_p\left( \|\bB \|^{t-1} \right), 	\\
	\left| L_T(\btheta, \bh_1) - L_T(\btheta, \bh_{0,1})\right|  &=  O_p\left(\frac{1}{T} \right).
\end{align*}
\end{lemma}
\textbf{ Proof of Lemma \ref{lemma:irr}.}
By the sub-multiplicativity, we have
\begin{align*}
	\left\| \bh_t(\btheta, \bh_1) - \bh_t(\btheta, \bh_{0,1} )\right\|  &=  \left\| \bB^{t-1}\bh_1 - \bB^{t-1}\bh_{0,1} \right\| 	\\
	 &\le  \|\bB\|^{t-1} \|\bh_1 - \bh_{0,1}\| 	\\
	 &=  C \|\bB \|^{t-1} .
\end{align*}
This implies that
$$ 
\left| L_T(\btheta, \bh_1) - L_T(\btheta, \bh_{0,1})\right|   =  O_p\left(\frac{1}{T} \right).
$$
\endpf

 Lemma \ref{lemma:irr} shows that the effect of the initial value is negligible. 
 Thus, for the mathematical convenience, we assume that $\bh_1(\btheta) = (\bI_r - \bB)^{-1} \bomega + \sum_{k=0}^\infty{\bB^k \bA \bff_{-k}^2}$.
Then we have $\bh_t(\btheta) =(\bI_r-\bB)^{-1}\bomega + \sum_{k=0}^\infty{\bB^k \bA \bff_{t-1-k}^2}$. 
To obtain the consistent $\hat{\btheta}$, it is sufficient to show that $\hat{L}_T(\btheta) \xrightarrow{p} L_{0,T}$ uniformly in $\btheta$ (Lemma \ref{lemma:unif_conv}) and $L_{0,T}(\btheta)$ attains a unique global maximum at $\btheta_0$ \citep{amemiya1985advanced}.

\begin{lemma}[Uniform Convergence] 	\label{lemma:unif_conv}
Under the assumptions of Theorem \ref{thm:GARCH}, we have
\begin{align*}
	\hat{L}_T(\btheta) \xrightarrow{p} L_{0,T}(\btheta)\text{ uniformly in }\btheta.
\end{align*}
\end{lemma}

\textbf{Proof of Lemma \ref{lemma:unif_conv}.} 
We have
\begin{align*}
	| \hat{L}_T(\btheta) - L_{0,T}(\btheta) |  \leq | \hat{L}_T(\btheta) - L_T(\btheta)| + |L_T(\btheta) - L_{0,T}(\btheta)|.
\end{align*}
For the simplicity, we omit the parameter $\btheta$.
Consider $ | \hat{L}_T(\btheta) - L_T(\btheta)|$. 
We have
\begin{align*}
	\left|\hat{L}_T(\btheta) - L_T(\btheta) \right| &\le \frac{1}{T}\sum_{t=1}^T{\left(\Big| \sum_{i=1}^r{\log{\hat{h}_{it}} - \log{h_{it}}} \Big| + \big| \hat{\bff}_t^2 - \bff_t^2\big|^\T\hat{\bh}_t^{-1} + {\bff_t^2}^\T \big|\hat{\bh}_t^{-1} - \bh_t^{-1} \big|\right)} 	\\
	&=\frac{1}{T}\sum_{t=1}^T{\Big(\text{(I)}+\text{(II)}+\text{(III)} \Big)},
\end{align*}
where $\hat{\bh}_t^{-1}$ is the element-wise inverse of $\hat{\bh}_t$. 
Recall $\bh_t(\btheta) =(\bI_r-\bB)^{-1}\bomega + \sum_{k=0}^\infty{\bB^k \bA \bff_{t-1-k}^2}$. 
Then by the fact that $\log{(1+x)}\le x$ for all $x>-1$, we have
\begin{align*}
	\sup_{\btheta} \text{(I)} &= \sup_{\btheta} \left|\sum_{i=1}^r{\log{\left(1+\frac{\Delta \hat{h}_{it}}{h_{it}}\right)}}\right| 	\\
		&\le \sup_{\btheta}\sum_{i=1}^r{\left|\frac{\Delta \hat{h}_{it}}{h_{it}}\right|} 	\\
		&\le \frac{1}{\min_{i\le r}{\omega_{\min,i}}} \1^\T \sup_{\btheta}\left| \sum_{k=0}^\infty{\bB^k \bA \big(\Delta\hat{\bff}_{t-1-k}^2\big)}\right| 	\\
		&= o_p(1),
\end{align*}
where $\Delta\hat{\bh}_t=\hat{\bh}_t - \bh_t$, $\Delta\hat{\bff}_t^2=\hat{\bff}_t^2 - \bff_t^2$, and the last equality is due to  Theorem \ref{thm:factor}. 
Similarly, we can show that $\text{(II)}$ and $\text{(III)}$ uniformly converges to zero. 
Therefore, we have
\begin{equation*} 
 \sup_{\btheta} | \hat{L}_T(\btheta) - L_T(\btheta)| = o_p(1).
\end{equation*}

Now we consider $L_T(\btheta) - L_{0,T}(\btheta)$. 
By Theorem 2.1 in \citet{newey1991uniform}, uniform convergence is equivalent to the pointwise convergence and stochastic equicontinuity, where the stochastic equicontinuity is satisfied by the Lipschitz condition $\big|\partial_\theta\big(L_T(\btheta)-L_{0,T}(\btheta)\big)\big|\le O_p(1)$. 
Thus, it is enough to show 
\begin{enumerate}
\item $L_T(\btheta) \xrightarrow{p}L_{0,T}(\btheta)$ pointwise in $\btheta$;
\item $\bbE\big[\sup_{\btheta}\big|\partial_{\theta} \big(L_T(\btheta) - L_{0,T}(\btheta)\big) \big| \big] < \infty$ for all $\theta \in \btheta$.
\end{enumerate}
Since $\bff_t^2 -\bh_{0,t}$ is a martingale difference, we can show 
\begin{align*}
	\bbE\left[\left( L_T(\btheta) - L_{0,T}(\btheta)\right)^2 \right] = o(1).
\end{align*}
Now, we show the Lipschitz condition.
We have
\begin{align*}
\bbE\big[\sup_{\btheta}\big|\partial_{\theta} \big(L_T(\btheta) - L_{0,T}(\btheta)\big) \big| \big]  &=  \bbE\bigg[\sup_{\btheta}\bigg| \frac{1}{T}\sum_{t=1}^T{\sum_{i=1}^{r}{ - h_{it}^{-2} (f_{it}^2 - h_{0,it})\partial_{\theta}h_{it} }} \bigg| \bigg] 	\\
	&\le C \frac{1}{T} \sum_{t=1}^T{\sum_{i=1}^r\bbE\left[\sup_{\btheta}(\partial_\theta h_{it})\left( f_{it}^{2} + h_{0,it} \right) \right]} 	\\
	&=  \frac{C}{T} \sum_{t=1}^T{\sum_{i=1}^r\bbE\left[2\sup_{\btheta}(\partial_\theta h_{it})\left( h_{0,it} \right)\right]} 	\\
	&\le  \frac{C}{T} \sum_{t=1}^T{\sum_{i=1}^r{\bbE\left[ h_{0,it}^2\right]^{\frac{1}{2}} \bbE \left[\sup_{\btheta}(\partial_\theta h_{it})^2\right]^{\frac{1}{2}}}} 	\\
	&< \infty,
\end{align*}
where the first inequality is due to $\partial_{\theta}h_{it} > 0$ for all $\btheta \in \bTheta$, and the second inequality is due to H\"older's inequality.
\endpf

\begin{lemma}[Uniqueness of $\btheta$] 	\label{lemma:glob_max}
Under the assumptions of Theorem \ref{thm:GARCH},
$$
\btheta^* = \arg \max_{\btheta}{L_{0,T}(\btheta)}
$$
 is unique almost surely, and $\btheta^* = \btheta_0$.
 Moreover,  we have
\begin{align} 	\label{eq:consistency}
	\hat{\btheta}\xrightarrow{p} \btheta_0.
\end{align}
\end{lemma}
\textbf{Proof of Lemma \ref{lemma:glob_max}.}
Since $\log{x} + t/x$ has a unique minimizer at $x=t$, $\bh_{it}(\btheta_0)$ is a unique maximizer of $l_{0,t}(\bh_{it})$ for all $t$. 
If $\btheta_0$ is not a unique parameter to have $\bh_{it}(\btheta_0)$, then there exists $\btheta^*\neq \btheta_0$ such that $\bh_t(\btheta^*) = \bh_t(\btheta_0)$. 
Then $\{\bh_t(\btheta^*) - \bh_t(\btheta_0)\}_{t\le T} = \{\0\}_{t\le T}$, and 
\begin{align*}
	&\begin{pmatrix}
		\bh_2(\btheta^*) - \bh_2(\btheta_0), \ldots, \bh_T(\btheta^*) - \bh_T(\btheta_0)
	\end{pmatrix} 	\\
	&= \begin{pmatrix}
		\bomega^* - \bomega_0 & \bA^* - \bA_0 & \bB^* - \bB_0
	\end{pmatrix}
	 \begin{pmatrix}
		1 & 1 & \cdots & 1 \\
		\bff_1^{2} & \bff_2^{2} & \cdots & \bff_{T-1}^{2} \\
		\bh_1(\btheta_0) & \bh_2(\btheta_0) & \cdots & \bh_{T-1}(\btheta_0)
	\end{pmatrix} \\
	&=\begin{pmatrix}
		\bomega^* - \bomega_0 & \bA^* - \bA_0 & \bB^* - \bB_0
	\end{pmatrix} \bM  =\B0,
\end{align*}
where $\B0$ is a zero matrix. 
By Assumption \ref{ass:GARCH}(b),   $\bM\bM^\T$ is invertible a.s., which implies 
$$\begin{pmatrix}
\bomega^* - \bomega_0 & \bA^* - \bA_0 & \bB^* - \bB_0 \end{pmatrix} = \B0.
$$ 
Therefore, $\btheta^* = \btheta_0$ $\text{a.s.}$ which is a contradiction.

With the uniqueness solution result, by Lemma  \ref{lemma:unif_conv}  and Theorem 4.1.2 in \citet{amemiya1985advanced}, we can show  \eqref{eq:consistency}.
\endpf

\begin{lemma} 	\label{lemma:bound_del}
Under the assumptions of Theorem \ref{thm:GARCH}, for all fixed $c \ge 1$, $l \le r$ and $\theta \in \btheta$, 
$$
	\bbE\left[\left|\sup_{\btheta}{\frac{\partial_\theta \hat{h}_{lt}}{\hat{h}_{lt}}} \right|^c \right] < \infty, \; \bbE\left[\left|\sup_{\btheta}{\frac{\partial_\theta^2 \hat{h}_{lt}}{\hat{h}_{lt}}} \right|^c \right] < \infty, \; \bbE\left[\left|\sup_{\btheta}{\frac{\partial_\theta^3 \hat{h}_{lt}}{\hat{h}_{lt}}} \right|^c \right] < \infty.
$$
\end{lemma}
\textbf{Proof of Lemma \ref{lemma:bound_del}.}
Notice that
$$
	\partial_\theta \hat{\bh}_t = \partial_\theta \bvarpi + \sum_{k=0}^\infty{\bB^k (\partial_\theta \bA) \hat{\bff}_{t-1-k}^{2}} + \sum_{k=1}^\infty{\left(\sum_{\xi=0}^{k-1} {\bB^\xi (\partial_\theta \bB) \bB^{k-1-\xi}}\right) \bA \hat{\bff}_{t-1-k}^{2}} \ge \0.
$$
where $\bvarpi=(\bI_r-\bB)^{-1}\bomega$.
For the case of $\theta = \varpi_i$, $\bbE\big[\big|\sup_{\btheta}{ \partial_\theta \hat{h}_{lt} / \hat{h}_{lt}} \big|^c \big] < \infty$ is trivial. 
With the case $\theta=A_{ij}$, 
\begin{align*}
\frac{\partial_\theta \hat{h}_{lt}}{\hat{h}_{lt}} &\le \frac{\be_l^\T \sum_{k=0}^\infty{\bB^k(\partial_\theta \bA )\hat{\bff}_{t-1-k}^{2} }}{\be_l^\T \sum_{k=0}^\infty{\bB^k \bA \hat{\bff}_{t-1-k}^{2}}} \\
&\le \frac{\be_l^\T  \sum_{k=0}^\infty{\bB^k \be_i \be_j^\T \hat{\bff}_{t-1-k}^{2} }}{\be_l^\T \sum_{k=0}^\infty{\bB^k  \be_i  A_{ij} \be_j^\T \hat{\bff}_{t-1-k}^{2}}} \\
&=\frac{1}{A_{ij}} \le  \frac{1}{A_{\min,ij}},
\end{align*}
where $\be_i$ is the $i^{th}$ standard basis vector.
Thus,  $\bbE\big[\big|\sup_{\btheta}{ \partial_\theta \hat{h}_{lt} / \hat{h}_{lt}} \big|^c \big] < \infty$.
For the case of $\theta = B_{ij}$, we have
\begin{align*}
\frac{\partial_\theta \hat{h}_{lt}}{\hat{h}_{lt}} &= \frac{\be_l^\T \sum_{k=1}^\infty{\left(\sum_{\xi=0}^{k-1} {\bB^\xi (\partial_\theta \bB) \bB^{k-1-\xi}}\right) \bA \hat{\bff}_{t-1-k}^{2}}}{\varpi_l + \be_l^\T \sum_{k=0}^\infty{\bB^k \bA \hat{\bff}_{t-1-k}^{2}}} 	\\
&\le  \sum_{k=1}^\infty{ \frac{ \be_l^\T \left(\sum_{\xi=0}^{k-1} {  \bB^\xi \be_i  \be^\T_j \bB^{k-1-\xi}}\right) \bA \hat{\bff}_{t-1-k}^{2} }{ \varpi_l +  \be_l^\T \left( \sum_{\xi=0}^{k-1}\bB^\xi\be_i B_{ij}  \be^\T_j \bB^{k-1-\xi}\right) \bA \hat{\bff}_{t-1-k}^{2}} } 	\\
&=  \sum_{k=0}^\infty{\frac{ \be_l^\T \left(\sum_{\xi=0}^{k-1} {  \bB^\xi \be_i  \be^\T_j \bB^{k-1-\xi}}\right) \bA \hat{\bff}_{t-1-k}^{2} }{ \varpi_l + B_{ij} \be_l^\T \left(\sum_{\xi=0}^{k-1} {  \bB^\xi \be_i  \be^\T_j \bB^{k-1-\xi}}\right) \bA \hat{\bff}_{t-1-k}^{2}}} 	\\
&=  \frac{1}{B_{ij}}\sum_{k=0}^\infty{\frac{(B_{ij}/\varpi_l) \be_l^\T \left(\sum_{\xi=0}^{k-1} {  \bB^\xi \be_i  \be^\T_j \bB^{k-1-\xi}}\right) \bA \hat{\bff}_{t-1-k}^{2} }{1 + (B_{ij}/\varpi_l) \be_l^\T \left(\sum_{\xi=0}^{k-1} {  \bB^\xi \be_i  \be^\T_j \bB^{k-1-\xi}}\right) \bA \hat{\bff}_{t-1-k}^{2}}} 	\\
&\le \frac{1}{B_{ij}}\frac{B_{ij}^\alpha}{\varpi_l^\alpha} \sum_{k=0}^\infty{\left[ \be_l^\T \left(\sum_{\xi=0}^{k-1} {  \bB^\xi \be_i  \be^\T_j \bB^{k-1-\xi}}\right) \bA \hat{\bff}_{t-1-k}^{2}\right]^\alpha }  \\
&\le \frac{1}{B_{ij}}\frac{1}{\varpi_l^\alpha} \sum_{k=0}^\infty{\left[\be_l^\T\bB^k \bA \hat{\bff}_{t-1-k}^{2}\right]^\alpha } ,
\end{align*}
where the second inequality is due to $x/(1+x)\le x^\alpha$ for $\forall{\alpha}\in[0,1]$. 
By choosing $\alpha = 1/c$, we have
\begin{align*}
\left(\bbE\left[\sup_{\btheta}\left|\frac{\partial_\theta \hat{h}_{lt}}{\hat{h}_{lt}}\right|^c\right]\right)^{\frac{1}{c}} &\le C\sum_{k=0}^\infty \|\bB_{\max} \|^k  \bbE\[ \| \hat{\bff}_{t-1-k}^{2} \| \]	\\
& < \infty,
\end{align*}
where first inequality is due to the Minkowski's inequality, and the last inequality follows from $\bbE\big[|f_{it}|^8\big]\le C$ and $\bbE\big\|\hat{\bff}_t^2 - \bff_t^2\big\|^2 = o(1)$.
Similarly, we can show the higher order derivatives bounds.
\endpf

\begin{lemma} 	\label{lemma:conv_rate}
Under the assumptions of Theorem \ref{thm:GARCH}, let $\bbE\left\|\Delta \bff_t^{2} \right\|^2 = O(\beta_T^2)$. 
Then we have
$$
	\left|\partial_\theta \hat{L}_T(\btheta_0) - \partial_\theta L_{0,T}(\btheta_0)\right| = O_p\left(\beta_T + \frac{1}{\sqrt{T}}\right).
$$
\end{lemma}
\textbf{Proof of Lemma \ref{lemma:conv_rate}.}
Denote $\hat{\bh}_{0,t} = \hat{\bh}_t(\btheta_0)$ and $\partial_{\theta}\hat{\bh}_{0,t} = \partial_{\theta}\hat{\bh}_t(\btheta_0)$.
Since $\bbE\left\|\Delta \bff_t^{2} \right\|^2 = O(\beta_T^2)$, we have
\begin{enumerate}
\item $\bbE\left\|\hat{\bh}_{0,t} - \bh_{0,t} \right\|^2 = O(\beta_T^2)$;
\item $\bbE\left\|\hat{\bh}_{0,t}^{-1} - \bh_{0,t}^{-1} \right\|^2 = O(\beta_T^2)$;
\item $\bbE\left\|\partial_\theta\hat{\bh}_{0,t} - \partial_\theta\bh_{0,t} \right\|^2 = O(\beta_T^2)$;
\item $\bbE\left[ {\bff_t^{2}}^\T(\hat{\bh}_{0,t}^{-2}-\bh_{0,t}^{-2}) \circ \partial_\theta \hat{\bh}_{0,t} \right] = O(\beta_T)$,
\end{enumerate}  
where the last equation is by the fact that
\begin{align*}
&\bbE\left[ {\bff_t^{2}}^\T(\hat{\bh}_{0,t}^{-2}-\bh_{0,t}^{-2}) \circ \partial_\theta \hat{\bh}_{0,t} \right]  	\\
	&= \bbE\left[\sum_{l=1}^r{\left(\frac{h_{0,lt}^{2}-\hat{h}_{0,lt}^{2}}{h_{0,lt}^{2}\hat{h}_{0,lt}^{2}}\right)f_{lt}^{2} \partial_\theta \hat{h}_{0,lt}}\right]	 \\
	&= \bbE\left[\sum_{l=1}^r{\left(\frac{h_{0,lt}^{2}-\hat{h}_{0,lt}^{2}}{h_{0,lt}\hat{h}_{0,lt}}\right) \epsilon_{lt}^{2} \frac{\partial_\theta \hat{h}_{0,lt}}{\hat{h}_{0,lt}}}\right]	 \\
	&= \bbE\left[\sum_{l=1}^r{\left(\frac{h_{0,lt}-\hat{h}_{0,lt}}{\hat{h}_{0,lt}}\right) \epsilon_{lt}^{2} \frac{\partial_\theta \hat{h}_{0,lt}}{\hat{h}_{0,lt}} + \left(\frac{h_{0,lt}-\hat{h}_{0,lt}}{h_{0,lt}}\right) \epsilon_{lt}^{2} \frac{\partial_\theta \hat{h}_{0,lt}}{\hat{h}_{0,lt}}}\right] 	\\
  &\le C \bbE\left[\sum_{l=1}^r{\left|h_{0,lt}-\hat{h}_{0,lt}\right| \epsilon_{lt}^{2} \frac{\partial_\theta \hat{h}_{0,lt}}{\hat{h}_{0,lt}}}\right] 	\\
	&\le C \sum_{l=1}^r{ \left( \bbE\left[\left(h_{0,lt}-\hat{h}_{0,lt}\right)^2\right] \right)^{\frac{1}{2}} \left( \bbE\left[\left(\frac{\partial_\theta \hat{h}_{0,lt}}{\hat{h}_{0,lt}}\right)^2\right]\right)^{\frac{1}{2}}} \\
	&= O(\beta_T),
\end{align*}
where the last equality is due to Lemma \ref{lemma:bound_del}. 
Thus, we have
\begin{align*}
&\partial_\theta \hat{L}_T (\btheta_0) - \partial_\theta L_T(\btheta_0) 	\\
	&= - \frac{1}{T}\sum_{t=1}^T{ (\hat{\bh}_{0,t} - \hat{\bff}_t^{2})^\T(\hat{\bh}_{0,t}^{-2} \circ \partial_\theta \hat{\bh}_{0,t}) - (\bh_{0,t} - \bff_t^{2})^\T(\bh_{0,t}^{-2} \circ \partial_\theta \bh_{0,t})} \\
	&= \frac{1}{T}\sum_{t=1}^T\bigg\{(\hat{\bff}_{t}^{2} - \bff_t^{2})^\T(\hat{\bh}_{0,t}^{-2} \circ \partial_\theta \hat{\bh}_{0,t}) + {\bff_t^{2}}^\T(\hat{\bh}_{0,t}^{-2}-\bh_{0,t}^{-2}) \circ \partial_\theta \hat{\bh}_{0,t} \\
	&\qquad\qquad + {\bff_t^{2}}^\T \bh_{0,t}^{-2} \circ (\partial_\theta \hat{\bh}_{0,t} - \partial_\theta \bh_{0,t} ) - (\hat{\bh}_{0,t}^{-1} - \bh_{0,t}^{-1})^\T\partial_\theta \hat{\bh}_{0,t} \\
	&\qquad\qquad - {\bh_{0,t}^{-1}}^\T (\partial_\theta \hat{\bh}_{0,t} - \partial_\theta \bh_{0,t}) \bigg\}
\end{align*}
and
\begin{align} 	\label{eq:hatL_T-L_T}
\bbE\left|\partial_\theta \hat{L}_T (\btheta_0) - \partial_\theta L_T(\btheta_0) \right| &\le \frac{C}{T}\sum_{t=1}^T\left\{ \beta_T + 4\beta_T  + \beta_T +  \beta_T + \beta_T \right\} 	\nonumber \\
	&=  O(\beta_T). 
\end{align}
Moreover, we have
\begin{align*}
	\partial_\theta L_T (\btheta_0) - \partial_\theta L_{0,T}(\btheta_0) &= \frac{1}{T}\sum_{t=1}^T{(\bff_t^{2} - \bh_{0,t})^\T(\bh_{0,t}^{-2} \circ \partial_\theta \bh_{0,t})} \\
	&= \frac{1}{T}\sum_{t=1}^T{(\beps_t^{2} - \1)^\T(\bh_{0,t}^{-1} \circ \partial_\theta \bh_{0,t})}
\end{align*}
and
\begin{align} 	\label{eq:L_T-L_0t}
&\bbE[(\partial_\theta L_T (\btheta_0) - \partial_\theta L_{0,T}(\btheta_0))^2]  \nonumber \\
	&= \frac{1}{T^2}\sum_{t=1}^T{\tr{\left((\bbE[\beps_t^{2}{\beps_t^{2}}^\T] - \1 \1^\T)\bbE\left[ (\bh_{0,t}^{-1} \circ \partial_\theta \bh_{0,t})(\bh_{0,t}^{-1}  \circ \partial_\theta \bh_{0,t})^\T\right] \right)}} \nonumber \\
	&= O\left(\frac{1}{T} \right),
\end{align}
where the last equality is due to Lemma \ref{lemma:bound_del}.
By combining \eqref{eq:hatL_T-L_T} and \eqref{eq:L_T-L_0t}, we complete the proof.
\endpf


\begin{lemma} 	\label{lemma:jacob_pd} 
Under the assumptions of Theorem \ref{thm:GARCH}, we have
$$
	\partial_{\btheta}^2 \hat{L}_T(\btheta^*) \xrightarrow{p} \partial_{\btheta}^2 L_{0,T}(\btheta_0),
$$
and $-\partial_{\btheta}^2 L_{0,T}(\btheta_0)$ is a positive definite matrix.
\end{lemma}
\textbf{Proof of Lemma \ref{lemma:jacob_pd}.}
We have
\begin{align*}
\partial_{\btheta}^2\hat{L}_T(\btheta^*) - \partial_{\btheta}^2 L_{0,T}(\btheta_0) &= \left(\partial_{\btheta}^2\hat{L}_T(\btheta^*) - \partial_{\btheta}^2 \hat{L}_{T}(\btheta_0)\right) + \left(\partial_{\btheta}^2\hat{L}_T(\btheta_0) - \partial_{\btheta}^2 L_{0,T}(\btheta_0)\right) 	\\
&= \text{(I)} + \text{(II)}.
\end{align*}
By the mean value theorem and \eqref{eq:consistency}, we have $\text{(I)} = \partial_{\btheta}^3\hat{L}_T(\btheta^{**})(\btheta^* - \btheta_0)$ and $\btheta^*-\btheta_0 \le \big| \hat{\btheta} - \btheta_0\big| \xrightarrow{p} 0$, respectively.
Thus, it is enough to show $\partial_{\btheta}^3\hat{L}_T(\btheta^{**}) = O_p(1)$. Denote $\partial_{ijk}^3 = \partial_{\theta_i} \partial_{\theta_j} \partial_{\theta_j}$. 
Then, for all $i,j,k \le \dim(\btheta)$,  we have
\begin{align*}
&\partial_{ijk}^3\hat{L}_T(\btheta) 	\\
	&=  \frac{1}{T}\sum_{t=1}^T\bigg\{\left(\hat{\bh}_t - \hat{\bff}_t^{2}\right)^\T \hat{\bh}_t^{\circ -2} \circ \partial_{ijk}^3 \hat{\bh}_t \\
	&\qquad\qquad +\left(2\hat{\bff}_t^{2} - \hat{\bh}_t \right)^\T \hat{\bh}_t ^{-3} \circ \left(\partial_i \hat{\bh}_t \circ \partial_{jk}\hat{\bh}_t + \partial_j\hat{\bh}_t \circ \partial_{ki}\hat{\bh}_t + \partial_k \hat{\bh}_t \circ \partial_{ij} \hat{\bh}_t  \right) \\
	&\qquad\qquad + \left(2\hat{\bh}_t - 6\hat{\bff}_t^{2}\right)^\T\hat{\bh}_t^{-4} \circ \partial_i \hat{\bh}_t \circ \partial_j \hat{\bh}_t \circ \partial_k \hat{\bh}_t \bigg\}.
	\end{align*}
By $\bbE\big[\hat{\bff}_t\hat{\bff}_t^\T\big] < \infty$ and Lemma \ref{lemma:bound_del}, we can show $\bbE\big|\partial_{ijk}^3\hat{L}_T(\theta^{**})\big| < \infty$. 
Thus, $\text{(I)} \xrightarrow{p} 0$.
Similar to the proof of Lemma \ref{lemma:conv_rate}, we can show
\begin{align*}
\text{(II)} &= \partial_{ij}^2\hat{L}_T(\btheta_0) - \partial_{ij}^2 L_{0,T}(\btheta_0) \\
&= \partial_{ij}^2\hat{L}_T(\btheta_0) - \partial_{ij}^2 L_T(\btheta_0) + \partial_{ij}^2 L_T(\btheta_0)  - \partial_{ij}^2 L_{0,T}(\btheta_0) \\
&= O_p(\beta_T) + O_p\left(\frac{1}{\sqrt{T}}\right).
\end{align*}
Moreover, since $\bh_{t}^{-2} > \0$ and $\bff_t^2$'s are non-degenerate, we can show
$$
	-\partial_{\btheta}^2 L_{0,T}(\btheta_0) = \frac{1}{T}\sum_{t=1}^T\sum_{i=1}^r{{h_{0,it}^{-2}}\left(\partial_{\btheta} h_{0,it}\ {\partial_{\btheta} h_{0,it}}^\T\right)} \succ 0.
$$
\endpf

\textbf{Proof of Theorem \ref{thm:GARCH}}.
By the mean value theorem, there is  $\btheta^*$ between $\hat{\btheta}$ and $\btheta_0$ such that 
\begin{align*} 	 
	\hat{\btheta} - \btheta_0 &= \partial^2_{\btheta} \hat{L}_T(\btheta^*)^{-1}\left(\partial_{\btheta} \hat{L}_T(\hat{\btheta}) - \partial_{\btheta} \hat{L}_T(\btheta_0) \right) 	\nonumber \\
	 &= \partial^2_{\btheta} \hat{L}_T(\btheta^*)^{-1}\left(\partial_{\btheta} L_{0,T}(\btheta_0) - \partial_{\btheta}\hat{L}_T(\btheta_0) \right),
\end{align*}
where the last equality is by the fact that $\partial_{\btheta} \hat{L}_T(\hat{\btheta}) =  \partial_{\btheta} L_{0,T}(\btheta_0) = 0$. 
Then, by  Lemmas \ref{lemma:conv_rate} and \ref{lemma:jacob_pd}, we have
\begin{align*}
	\hat{\btheta} - \btheta_0 = O_p\left(\beta_T + \frac{1}{\sqrt{T}} \right),
\end{align*}
and thus, the result of Theorem \ref{thm:factor} completes the proof.\endpf

\section{Proof of Theorem \ref{thm:prediction}} 	\label{proof:thm:prediction}

\begin{assumption}     \label{ass:prediction} ~
\begin{enumerate}
\item [(a)] \label{ass:prediction:subgaussian}  The sample covariance estimator satisfies the concentration inequality, for any given $a>0$, 
$$
\mathbb{P} \left\{ \max_{ij} \Big| \hat{\bar{ \Sigma}}_{ij} - \bar{\Sigma}_{ij} \Big| \geq C_a \sqrt{ \frac{\log p }{T} } \right\} \leq p^{-a},
$$
where $C_a$ is a constant only depending on $a$.
\item [(b)] \label{ass:prediction:coherence} There is a constant $C$ such that
$$
     \frac{1}{r}\max_{i\le p} \sum_{j=1}^r V_{ij}^2 \le C.
$$
\item [(c)] \label{ass:prediction:away_zero} The smallest eigenvalue of $\bSigma_{u}$ stays away from $0$, and $\big| \Sigma_{u,ij} \big| \le C$ for all $i$, $j$.
\item [(d)] $p=o(T^2)$.
\end{enumerate}
\end{assumption}

\begin{remark}
Assumption \ref{ass:prediction}(a) is the sub-Gaussian condition. 
When investigating the high-dimensional inferences, the sub-Gaussian condition is essential.
Recently, several robust variance estimation methods have been developed, which can satisfy the sub-Gaussian condition Assumption \ref{ass:prediction}(a)  under the finite forth moment condition \citep{catoni2012challenging, fan2018robust, minsker2018sub, shin2021adaptive}.
Thus, the sub-Gaussian condition Assumption \ref{ass:prediction}(a) is not restrictive. 
On the other hand, we can investigate the asymptotic behaviors under  sub-Weilbull conditions \citep{kuchibhotla2018moving, vladimirova2020sub}.  
\end{remark}

\begin{remark}
Assumption \ref{ass:prediction}(b) is called the incoherence condition, which is widely assumed in analyzing the low-rank matrix \citep{candes2009exact,fan2017l}.
The basic intuition is that the factor loading matrix $\bV$ is not to be sparse.
That is, the factor affects almost all the stock returns.
Thus, under the factor model, the incoherence condition is acceptable. 
\end{remark}

\begin{lemma} 	\label{lemma:relative_factor_cov_bound}
Under the assumptions of Theorem \ref{thm:prediction}, we have
\begin{align} 	\label{eq:relative_factor_cov_bound}
	\left\| \hat{\bV}\hat{\bSigma}_{f,t+1} \hat{\bV}^\T - \bV \bSigma_{f,t+1} \bV^\T \right\|_{\bSigma_{t+1}} = O_p\left(\frac{1}{\sqrt{T}} +\frac{\sqrt{p}}{T} + \frac{s_p^2}{p\sqrt{p}} + \frac{s_p}{p}\right).
\end{align}
\end{lemma}
\textbf{Proof of Lemma \ref{lemma:relative_factor_cov_bound}.}
Without loss of generality, we assume that $\bO$ is the identity matrix. 
Similar to the proofs of Theorem 4.2 in \citet{kim2018large}, we have
\begin{align*}
	&\left\| \hat{\bV}\hat{\bSigma}_{f,t+1} \hat{\bV}^\T - \bV \bSigma_{f,t+1} \bV^\T \right\|_{\bSigma_{t+1}} 	\\
	&\le  \left\| \bV \bSigma_{f,t+1}^{\frac{1}{2}}\left(\bSigma_{f,t+1}^{-\frac{1}{2}}\hat{\bSigma}_{f,t+1} \bSigma_{f,t+1}^{-\frac{1}{2}} - \bI_r\right) \bSigma_{f,t+1}^{\frac{1}{2}} \bV^\T \right\|_{\bSigma_{t+1}} 	\\
	&\quad + \left\| (\hat{\bV} - \bV ) \hat{\bSigma}_{f,t+1} (\hat{\bV} - \bV )^\T  \right\|_{\bSigma_{t+1}} + 2\left\| \bV \hat{\bSigma}_{f,t+1} (\hat{\bV} - \bV )^\T  \right\|_{\bSigma_{t+1}} 	\\
	&= \text{(I)} + \text{(II)} + \text{(III)}.
\end{align*}
First consider $\text{(I)}$.
By the Sherman-Morrison-Woodbury formula \citep{higham2002accuracy}, we have
\begin{align*}
	(\bV \bSigma_{f,t+1}\bV^\T + \bSigma_u)^{-1} = \bSigma_u^{-1} - \bSigma_u^{-1} \bV\bSigma_{f,t+1}^{\frac{1}{2}} (\bI_r + \bSigma_{f,t+1}^{\frac{1}{2}}\bV^\T \bSigma_u^{-1} \bV \bSigma_{f,t+1}^{\frac{1}{2}} )^{-1} \bSigma_{f,t+1}^{\frac{1}{2}} \bV^\T  \bSigma_u^{-1}
\end{align*}
and by denoting $\bX = \bSigma_{f,t+1}^{\frac{1}{2}} \bV^\T \bSigma_{u}^{-1} \bV \bSigma_{f,t+1}^{\frac{1}{2}}$,
\begin{align*}
	\left\|\bSigma_{f,t+1}^{\frac{1}{2}} \bV^\T \bSigma_{t+1}^{-1} \bV \bSigma_{f,t+1}^{\frac{1}{2}}\right\| 	&= \left\|\bX - \bX (\bI_r + \bX)^{-1} \bX \right\|  	\\
	&= \left\|\bX (\bI_r + \bX)^{-1} \right\|  	\\
	&= \left\|\bI_r - (\bI_r + \bX)^{-1} \right\| 	\\
	&\le 2.
\end{align*}
Then, since $\|\bA\bB\|_F \le \|\bA\| \|\bB\|_F$ and $\big\|\bSigma_{f,t+1}^{\frac{1}{2}} \bV^\T \bSigma_{t+1}^{-\frac{1}{2}}\big\| \le \sqrt{2}$, we have 
\begin{align*}
	\text{(I)} & =  \frac{1}{\sqrt{p}}\left\|\bSigma_{t+1}^{-\frac{1}{2}} \bV \bSigma_{f,t+1}^{\frac{1}{2}}\left(\bSigma_{f,t+1}^{-\frac{1}{2}} \hat{\bSigma}_{f,t+1} \bSigma_{f,t+1}^{-\frac{1}{2}} - \bI_r\right)\bSigma_{f,t+1}^{\frac{1}{2}}\bV^\T \bSigma_{t+1}^{-\frac{1}{2}} \right\|_F 	\\
	&\le  \frac{2}{\sqrt{p}}  \left\| \bSigma_{f,t+1}^{-\frac{1}{2}} \hat{\bSigma}_{f,t+1}\bSigma_{f,t+1}^{-\frac{1}{2}}  - \bI_r  \right\|_F 	\\
	&=  \frac{2}{\sqrt{p}}  \left(\sum_{i=1}{\frac{\left| \hat{h}_{it+1}(\hat{\btheta}) - h_{0,it+1}\right|^2}{h_{0,it+1}^2}}\right)^{\frac{1}{2}} 	\\
	&=  O_p\left(\frac{1}{\sqrt{pT}} + \frac{\sqrt{s_p}}{p} \right),
\end{align*}
where the last equality is by the fact that
\begin{align*}
	\hat{h}_{it+1}(\hat{\btheta}) - h_{0,it+1} &= \hat{h}_{it+1}(\hat{\btheta}) - \hat{h}_{it+1}(\btheta_0) + \hat{h}_{it+1}(\btheta_0) - h_{0,it+1} 	\\
	&= O_p \left( \frac{1}{\sqrt{T}} + \sqrt{\frac{s_p}{p}}\right).
\end{align*}
For \text{(II)} and \text{(III)}, similarly, by Theorem \ref{thm:factor}, we can show
\begin{align*}
	\text{(II)}  &=  \frac{1}{\sqrt{p}}\left\| \bSigma_{t+1}^{-\frac{1}{2}} (\hat{\bV} - \bV ) \hat{\bSigma}_{f,t+1} (\hat{\bV} - \bV  )^\T \bSigma_{t+1}^{-\frac{1}{2}}\right\|_F 	\\
	&\le  \frac{1}{\sqrt{p}} \left\| \bSigma_{t+1}^{-1} \right\| \left\|\hat{\bSigma}_{f,t+1} \right\|_F \left\| \hat{\bV} - \bV   \right\|^2 	\\
	&=  O_p\left(\frac{\sqrt{p}}{T} + \frac{s_p^2}{p\sqrt{p}} \right)
\end{align*}
and
\begin{align*}
	\text{(III)} & =  \frac{1}{\sqrt{p}} \left\| \bSigma_{t+1}^{-\frac{1}{2}} \bV \hat{\bSigma}_{f,t+1} (\hat{\bV} - \bV )^\T \bSigma_{t+1}^{-\frac{1}{2}}\right\|_F 	\\
	&=  \frac{\sqrt{2}}{\sqrt{p}} \left\| \hat{\bSigma}_{f,t+1}^{\frac{1}{2}}(\hat{\bV} - \bV  )^\T  \bSigma_{t+1}^{-\frac{1}{2}} \right\|_F 	\\
	&\le  \frac{C}{\sqrt{p}} \left\| \hat{\bSigma}_{f,t+1}^{\frac{1}{2}} \right\|_F \left\| \hat{\bV} - \bV    \right\| 	\\
	&=  O_p\left(\frac{1}{\sqrt{T}} + \frac{s_p}{p} \right).
\end{align*}
\endpf

\begin{lemma} 	\label{lemma:threshold_bound}
Under the assumptions of Theorem \ref{thm:prediction}, 
suppose that $\|\hat{\bSigma}_u - \bSigma_u\|_{\max} = O_p(\gamma_T)$, and the thresholding level satisfies the condition $\tau_T = C \gamma_T$ such that $| \hat{\Sigma}_{u,ij} - \Sigma_{u,ij} | < \tau_T (\hat{\Sigma}_{u,ii} \hat{\Sigma}_{u,jj})^{1/2}/2 $.
Then we have
$$
	\left\| \cT(\hat{\bSigma}_u) - \bSigma_u \right\| =  O_p\left(s_p \gamma_T^{1-q}\right).
$$
\end{lemma}

\textbf{Proof of Lemma \ref{lemma:threshold_bound}.} 
Let $\tau_{ij} = \tau_T (\hat{\Sigma}_{u,ii} \hat{\Sigma}_{u,jj})^{1/2}$.
Similar to the proofs of Section 3 in \citet{fan2018robust}, under the event $| \hat{\Sigma}_{u,ij} - \Sigma_{u,ij} | < \tau_{ij}/2 $, we have
\begin{align*}
	\left\| \cT(\hat{\bSigma}_u) - \bSigma_u \right\| &\le \left\| \cT(\hat{\bSigma}_u) - \bSigma_u \right\|_\infty 	\\
	&=  \max_i\sum_{j=1}^p{\left|s_{ij}(\hat{\Sigma}_{u,ij})\ind_{|\hat{\Sigma}_{u,ij}|\ge \tau_{ij}} - \Sigma_{u,ij}\ind_{|\Sigma_{u,ij}|\ge \tau_{ij}} - \Sigma_{u,ij}\ind_{|\Sigma_{u,ij}|< \tau_{ij}}  \right|} \\
	&\le 	\max_i\sum_{j=1}^p\bigg\{\left|s_{ij}(\hat{\Sigma}_{u,ij}) - \Sigma_{u,ij}\right|\ind_{|\hat{\Sigma}_{u,ij}|\ge \tau_{ij}} + \left|\Sigma_{u,ij}\right|\left|\ind_{|\hat{\Sigma}_{u,ij}|\ge \tau_{ij}} - \ind_{|\Sigma_{u,ij}|\ge \tau_{ij}} \right| 	\\
	&\qquad\qquad\quad + \left|\Sigma_{u,ij}\right|\ind_{|\Sigma_{u,ij}|< \tau_{ij}}\bigg\} \\
	&\le 	\max_i\sum_{j=1}^p\bigg\{\frac{3}{2}\tau_{ij}\ind_{|\Sigma_{u,ij}| \ge \frac{1}{2}\tau_{ij}} + \left|\Sigma_{u,ij}\right|\ind_{|\Sigma_{u,ij}| \le \frac{3}{2}\tau_{ij}} + \left| \Sigma_{u,ij} \right|^q \tau_{ij}^{1-q}\bigg\} \\
	&\le 	C \max_i\sum_{j=1}^p{|\Sigma_{u,ij}|^q \tau_{ij}^{1-q}}  	\\
	&= O\left(s_p \gamma_T^{1-q}\right),
\end{align*}
where the last equality is due to the sparse condition.
\endpf

\textbf{Proof of Theorem \ref{thm:prediction}}. 
By Assumption \ref{ass:prediction:subgaussian}(a), we have
\begin{align} 	\label{eq:sample_var_max_error}
	\left\|\hat{\bar{\bSigma}} - \bar{\bSigma}\right\|_{\max} = O_p\left( \sqrt{\frac{\log{p }}{T}} \right).
\end{align}	 
We denote $\hat{\bv}_i$ and $\bv_i$  the $i^{th}$ column vectors of $\hat{\bV}$ and $\bV$, respectively.
Then by Theorem 2.1 in \citet{fan2017l},
\begin{align*}
	\sum_{i=1}^r{\left\|\hat{\bv}_i \hat{\bv}_i^\T - \bv_i \bv_i^\T\right\|_{\max}} 
	&\le  \sum_{i=1}^r{\left\{\left\| (\hat{\bv}_i - \bv_i)\hat{\bv}_i^\T \right\|_{\max} + \left\|\bv_i(\hat{\bv}_i-\bv_i)^\T\right\|_{\max}\right\}} 	\\
	&\le  \sum_{i=1}^r{\left\{\left\| \hat{\bv}_i - \bv_i \right\|_\infty \left\| \hat{\bv}_i - \bv_i + \bv_i \right\|_\infty + \left\| \hat{\bv}_i - \bv_i \right\|_\infty \left\| \bv_i \right\|_\infty\right\}} 	\\
	&\le  \sum_{i=1}^r{\left\{\left\| \hat{\bv}_i - \bv_i \right\|_\infty^2 + 2\left\| \hat{\bv}_i - \bv_i \right\|_\infty \left\| \bv_i \right\|_\infty\right\}} 	\\
	&\le  C \left( \frac{\left\| \hat{\bar{\bSigma}} - \bar{\bSigma} + \bSigma_u \right\|_\infty^2}{p^2\delta_{r}^2 } + \frac{\left\| \hat{\bar{\bSigma}} - \bar{\bSigma} + \bSigma_u \right\|_\infty}{p\delta_{r} } \right) 	 \nonumber\\
	&=  O_p\left(\sqrt{\frac{\log{p}}{T}} + \frac{s_p}{p} \right),
\end{align*}
where the last equality is due to $\|\bA\|_{\infty} \le p \|\bA\|_{\max}$.
Thus, we have
\begin{align} 	\label{eq:factor_max_error_bound}
	&\left\| \hat{\bV}\hat{\bSigma}_{f,t+1} \hat{\bV}^\T - \bV \bSigma_{f,t+1} \bV^\T \right\|_{\max} 	 \nonumber\\
	&\le  \sum_{i=1}^r{\left\{\left\|  \hat{h}_{it+1}\left(\hat{\bv}_i \hat{\bv}_i^\T - \bv_i \bv_i^\T\right)  \right\|_{\max}  +  \left\|  (\hat{h}_{it+1} - h_{0,it+1})\bv_i \bv_i^\T  \right\|_{\max} \right\}}  	 \nonumber\\
	&\le  \left\|  \hat{\bh}_{t+1} - \bh_{0,t+1} + \bh_{0,t+1}\right\|_\infty \sum_{i=1}^r{\left\|\hat{\bv}_i \hat{\bv}_i^\T - \bv_i \bv_i^\T\right\|_{\max}}  + \left\| \hat{\bh}_{t+1} - \bh_{0,t+1}\right\|_\infty \sum_{i=1}^r{ \left\|\bv_i \bv_i^\T  \right\|_{\max}} 	 \nonumber\\
	&\le O_p \left(\sqrt{\frac{\log{p}}{T}} + \frac{s_p}{p} \right) + O_p\left(\frac{1}{\sqrt{T}} + \sqrt{\frac{s_p}{p}}\right).
\end{align}
By combining \eqref{eq:sample_var_max_error} and  \eqref{eq:factor_max_error_bound}, we have
\begin{align} 	\label{eq:idio_max_error_bound}
	\left\|\hat{\bSigma}_u - \bSigma_u\right\|_{\max}  &\le  \left\| \hat{\bar{\bSigma}} - \bar{\bSigma} \right\|_{\max} + \left\| \hat{\bV}\hat{\bSigma}_{f,t+1} \hat{\bV}^\T - \bV \bSigma_{f,t+1} \bV^\T \right\|_{\max} 	 \nonumber\\
	&= O_p\left(\sqrt{\frac{\log{p}}{T}} + \sqrt{\frac{s_p}{p}} \right),
\end{align}
which leads to $\gamma_T = C\big(\sqrt{\log{p}/T} + \sqrt{s_p/p} \big)$ and $\|\hat{\bSigma}_{t+1} - \bSigma_{t+1}\|_{\max} = O_p(\gamma_T)$.
Combining Lemmas \ref{lemma:relative_factor_cov_bound} and \ref{lemma:threshold_bound}, and \eqref{eq:idio_max_error_bound}, we have
\begin{align*}
	\left\|\hat{\bSigma}_{t+1} - \bSigma_{t+1}\right\|_{\bSigma_{t+1}} &\le \left\|\hat{\bV}\hat{\bSigma}_{f,t+1} \hat{\bV}^\T - \bV \bSigma_{f,t+1} \bV^\T\right\|_{\bSigma_{t+1}} + \left\|\cT(\hat{\bSigma}_u) - \bSigma_u\right\|_{\bSigma_{t+1}} 	\\
	&=   O_p\left(\frac{\sqrt{p}}{T} + s_p \gamma_T^{1-q} \right),
\end{align*}
where the last equality is by the fact
\begin{align*}
	\left\|\cT(\hat{\bSigma}_u) - \bSigma_u\right\|_{\bSigma_{t+1}} &\le \frac{1}{\sqrt{p}}\left\|\cT(\hat{\bSigma}_u) - \bSigma_u\right\| \left\| \bSigma_{t+1}^{-1} \right\|_F 	\\
		&\le  \left\|\cT(\hat{\bSigma}_u) - \bSigma_u\right\| \left\| \bSigma_{t+1}^{-1} \right\| 	\\
		&= O_p\left(s_p \gamma_T^{1-q}\right).
\end{align*}
\endpf

\section{Proof of Theorem \ref{thm:prediction_VaR}} 	\label{proof:thm:prediction_VaR}

\begin{assumption}     \label{ass:prediction_VaR} $\,$
\begin{enumerate}
\item [(a)] \label{ass:prediction_VaR:subgaussian} The estimated standardized return $\hat{x}_t = \bw^\T(\by_t - \bar{\by})/(\bw^\T \hat{\bSigma}_t\bw)^{1/2}$ satisfies the concentration inequality, for any given $a>0$,
$$
\max_{t\le T} \mathbb{P} \left\{ \left| \hat{x}_t - x_t \right| \geq C_a \sqrt{ \log{T} } \left(\sqrt{\frac{\log{p}}{T}} + \sqrt{\frac{s_p}{p}}\right) \right\} \leq T^{-a},
$$
where $x_t=\bw^\T(\by_t - \bmu)/(\bw^\T \bSigma_t\bw)^{1/2}$ and $C_a$ is a constant only depending on $a$.
\item [(b)] \label{ass:prediction_VaR:pdf} The cumulative density function of $x_t$ and its inverse function are continuous. 

\item[(c)] The variance of the portfolio return $r_t$ is bounded and strictly positive.
\end{enumerate}
\end{assumption}

\begin{remark}
 The sub-Gaussian condition Assumption \ref{ass:prediction_VaR:subgaussian}(a) is true  under some sub-Gaussian condition for $\by_t$. 
 Assumption \ref{ass:prediction_VaR:pdf}(b) is usually imposed to study quantile estimations \citep{chen2005nonparametric}. 
\end{remark}

\textbf{Proof of Theorem \ref{thm:prediction_VaR}}.
Consider the parametric VaR estimator case.
By the definition of VaR in \eqref{eq:VaR}, we have
\begin{align*}
	\left|\hat{\text{VaR}}_{\alpha,t+1} - \text{VaR}_{\alpha,t+1}\right |  &\le C\left |c_{\alpha}\sqrt{\bw^\T \hat{\bSigma}_{t+1} \bw} - c_{\alpha}\sqrt{\bw^\T \bSigma_{t+1} \bw}\right| + C\left|\bw^\T(\bar{\by} -\bmu)\right|.
\end{align*}
Then, since $\|\bar{\by}-\bmu\|_{\max}^2=O_p(\log{p}/T)$ and $\|\bar{\by}-\bmu\|^2=O_p(p/T)$, we have
\begin{align} 	\label{eq:mean_conv}
	\left(\bw^\T(\bar{\by} -\bmu)\right)^2 = O_p\left(\min{\left\{\frac{\log{p}}{T}, \|\bw\|^2\frac{p}{T} \right\}}\right).
\end{align}
Consider $\Big|c_{\alpha}\sqrt{\bw^\T \hat{\bSigma}_{t+1} \bw} - c_{\alpha}\sqrt{\bw^\T \bSigma_{t+1} \bw}\Big|$. 
We have
\begin{align} 	\label{eq:var_conv_small_port}
	\left |c_{\alpha}\sqrt{\bw^\T \hat{\bSigma}_{t+1} \bw} - c_{\alpha}\sqrt{\bw^\T \bSigma_{t+1} \bw}\right | &\le C \left| \bw^\T \left(\hat{\bSigma}_{t+1} - \bSigma_{t+1}\right) \bw \right|  	\nonumber \\
	&\le C\sum_{i,j=1}^p{|w_i w_j|} \left\| \hat{\bSigma}_{t+1} - \bSigma_{t+1} \right\|_{\max}  	\nonumber \\
	&= O_p \left( \sqrt{\frac{\log{p}}{T}} + \sqrt{\frac{s_p}{p}} \right),
\end{align} 
where the last is due to Theorem \ref{thm:prediction} and $\|\bw\|_1 \le C$. 
Combining \eqref{eq:mean_conv} and \eqref{eq:var_conv_small_port}, we have
\begin{align*}
    \left |\hat{\text{VaR}}_{\alpha,t+1} - \text{VaR}_{\alpha,t+1}\right| = O_p\left( \sqrt{\frac{\log{p}}{T}} + \sqrt{\frac{s_p}{p}}\right).
\end{align*}

%
%
%

Consider the non-parametric $\sigma$-based VaR estimator case.
Define
\begin{align*}
	\hat{F}_T(x) = \frac{1}{T}\sum_{t=1}^T{\ind_{\{\hat{x}_t \le x\}}},\qquad F_T(x) = \frac{1}{T}\sum_{t=1}^T{\ind_{\{x_t \le x\}}},
\end{align*}
where $x$ is an $\alpha$-quantile value, and $F(x)$ as a cumulative distribution function of $x_t$.
Then the expected value of the absolute difference between $\hat{F}_T(x)$ and $F_T(x)$ is
\begin{align*}
	\bbE\left| \hat{F}_T(x) - F_T(x) \right| \le \frac{1}{T}\sum_{t=1}^T{\big( \bbP \{\hat{x}_t \le x, x_t > x \} + \bbP \{ \hat{x}_t > x, x_t \le x\} \big)}.
\end{align*}
Let $\vartheta_T = \sqrt{\log{p}/T} + \sqrt{s_p/p}$.
Then, since Assumption \ref{ass:prediction_VaR:subgaussian}(a) implies $|\hat{x}_t - x_t|=O_p(\vartheta_T\sqrt{ \log{T} })$, we have, for large enough $C$,
\begin{align*}
	\bbP \{\hat{x}_t \le x, x_t > x \} &\le \bbP \{ x_t \le x + |\hat{x}_t-x_t|, x_t > x \} 	\\
	&= \bbP \{ x < x_t \le x + |\hat{x}_t-x_t| \} 	\\
	&\le \bbP \{ x < x_t \le x + C \vartheta_T\sqrt{ \log{T} }\} + \bbP \{|\hat{x}_t-x_t| > C \vartheta_T\sqrt{ \log{T} }\} 	\\
	&\le F(x + C\vartheta_T \sqrt{ \log{T} }) - F(x) + \frac{1}{\sqrt{T}} 	\\
	&= O\left( \vartheta_T\sqrt{ \log{T} } + \frac{1}{\sqrt{T}} \right),
\end{align*}
where the last equality is due to Assumption \ref{ass:prediction_VaR:pdf}(b).
Similarly, the bound for $\bbP \{ \hat{x}_t > x, x_t \le x\}$ can be found.
Then $\big| \hat{F}_T(x) - F_T(x) \big|=O_p\big( \vartheta_T\sqrt{ \log{T} } + 1/\sqrt{T} \big)$.
By the Dvoretzky–Kiefer-Wolfowitz inequality \citep{dvoretzky1956asymptotic}, we have $\big| F_T(x) - F(x) \big|=O_p\big(1/\sqrt{T} \big)$.
Thus,
\begin{align*}
	\big| \hat{F}_T(x) - F(x) \big| &\le \big| \hat{F}_T(x) - F_T(x) \big|  + \big| F_T(x) - F(x) \big|  	\\
	&= O_p\big(\vartheta_T \sqrt{\log{T} }+ 1/\sqrt{T}\big).
\end{align*}
Define
\begin{align*}
	F^{-1}(y) = \inf{\big\{ x: F(x) > y \big\}}.
\end{align*}
 Then the $\lceil \alpha T \rceil$-th smallest value of $\{x_t\}_{t=1}^T$ is bounded as follows:
\begin{align*}
	\hat{F}^{-1}_T(\alpha) &= \inf{\big\{ x: \hat{F}_T(x) > \alpha \big\}} 	\\
	&\le  \inf{\big\{ x:  F(x) > \alpha  + \big| \hat{F}_T(x) - F(x) \big| \big\}} 	\\
	&\le F^{-1}\left(\alpha +   \big| \hat{F}_T(x) - F(x) \big| \right)
\end{align*}
and
\begin{align*}
	\hat{F}^{-1}_T(\alpha) &\ge  \inf{\big\{ x:  F(x) > \alpha  - \big| \hat{F}_T(x) - F(x) \big| \big\}} 	\\
	&\ge F^{-1}\left(\alpha - \big| \hat{F}_T(x) - F(x) \big|  \right).
\end{align*}
Thus, we have
\begin{align} 	\label{eq:sample_quant_conv}
	\big| \hat{F}_T^{-1}(\alpha) - F^{-1}(\alpha) \big| \le C  \big| \hat{F}_T(x) - F(x) \big|.
\end{align}
We also have
\begin{align} 	\label{eq:est_vol_bound}
	\bw^\T \hat{\bSigma}_{t+1} \bw &= \bw^\T \bSigma_{t+1} \bw  + \bw^\T (\hat{\bSigma}_{t+1} - \bSigma_{t+1}) \bw 	\nonumber \\
	&= \bw^\T \bV \bSigma_{f,t+1} \bV^\T \bw + \bw^\T \bSigma_u \bw + o_p(1) 	\nonumber \\
	&\le \max_{i\le r}{h_{it+1}} \left\| \bV^\T \bw \right\|^2 + C	+ o_p(1) 	\nonumber \\
	&= O_p(1),
\end{align}
where the inequality is due to the gross exposure condition and $|\Sigma_{u,ij}| \le C$, and the last equality is due to $\|\bV^\T \bw\|^2 = \|\sum_{i=1}^p{w_i\tilde{\bv}_i}\|^2 \le \sum_{i=1}^p{|w_i| \|\tilde{\bv}_i\|^2} \le C$, where $\tilde{\bv}_i$ is the $i^{th}$ row vector of $\bV$.
Combining results from \eqref{eq:mean_conv} to \eqref{eq:est_vol_bound}, we have
\begin{align*}
	\left |\hat{\text{VaR}}_{\alpha,t+1} - \text{VaR}_{\alpha,t+1}\right | &\le C \sqrt{\bw^\T \hat{\bSigma}_{t+1} \bw} \left | \hat{F}_T^{-1}(\alpha) - F^{-1}(\alpha) \right |  + C\left |\bw^\T(\bar{\by} -\bmu)\right | 	\\
	&\quad + C   F^{-1}(\alpha) \left |\sqrt{\bw^\T \hat{\bSigma}_{t+1} \bw} - \sqrt{\bw^\T \bSigma_{t+1} \bw}\right |	\\
	&= O_p\left(\vartheta_T \sqrt{ \log{T}}\right).
\end{align*}
\endpf

\bibliography{ref}

\end{spacing}
\end{document}